\documentclass{article}

\usepackage[english]{babel}

\usepackage{amsmath}
\usepackage{graphicx}
\usepackage[colorlinks=true, allcolors=blue]{hyperref}
\usepackage{enumerate}
\usepackage{float}
\usepackage{verbatim}
\usepackage{natbib}
\usepackage{amssymb}
\usepackage{algorithm2e}
\usepackage{verbatim}
\usepackage{dsfont}

\def \dsP {\text{$\mathds{P}$}}
\def \dsE {\text{$\mathds{E}$}}
\def \dsR {\text{$\mathds{R}$}}
\def \dsN {\text{$\mathds{N}$}}

    \def \mV {\text{\boldmath$V$}}

\usepackage{subcaption}
\usepackage{bm}
\usepackage{authblk}

\usepackage{amsthm}
\newtheorem{proposition}{Proposition}[section]
\theoremstyle{definition}
\newtheorem{remark}{Remark}

\newtheorem{definition}{Definition}

\newtheorem{theorem}{Theorem}

\title{Bayesian Effect Selection in Additive Models with an Application to Time-to-Event Data}

\author{Paul Bach and Nadja Klein}

\affil{\normalsize{Chair of Uncertainty Quantification and Statistical Learning\\ Research Center Trustworthy Data Science and Security (UA Ruhr) and \\Department of Statistics (Technische Universit\"at Dortmund)}\\
Joseph-von-Fraunhofer Straße 25, 44227 Dortmund}

\begin{document}

\def\spacingset#1{\renewcommand{\baselinestretch}%
{#1}\small\normalsize} \spacingset{1}

\maketitle

\begin{abstract}
Accurately selecting and estimating smooth functional effects in additive models with potentially many functions is a challenging task. We introduce a novel Demmler-Reinsch basis expansion to model the functional effects that allows us to orthogonally decompose an effect into its linear and nonlinear parts. We show that our representation allows to consistently estimate both parts as opposed to commonly employed mixed model representations. Equipping the reparameterized regression coefficients with normal beta prime spike and slab priors allows us to determine whether a continuous covariate has a linear, a nonlinear or no effect at all. We provide new theoretical results for the prior and  a compelling explanation for its superior Markov chain Monte Carlo mixing performance compared to the spike-and-slab group lasso. We establish an efficient posterior estimation scheme and illustrate our approach along effect selection on the hazard rate of a time-to-event response in  the geoadditive Cox regression model in  simulations and  data on survival with leukemia.

\end{abstract}\vspace{1em}
 \vspace{1em}

\newpage

\spacingset{1.9}

\section{Introduction}
Suppose we have data $\mathcal{D}=\{(y_i,\bm{x}_i),\ i=1,\dots,n\},$ where the $y_i$ are univariate response values and the $\bm{x}_i\in\dsR^p$ are vectors of covariates. A common model assumption is the generalized additive model \citep[GAM;][]{HasTib1986,HasTib1990}, where $Y\mid \bm{X}=(X_1,\dots,X_p)^T$ follows an exponential family distribution and the conditional mean is expressed as
\begin{align}\label{TrueGAM}
    g(\dsE[Y\mid \bm{X}])=\beta_0^\ast +\sum_{j=1}^p f^\ast_j(X_j),
\end{align}
where $g(\cdot)$ is a suitable link function and $\dsE[ f^\ast_j(X_j)]=0,\ j=1,\dots,p,$ to ensure identifiability. The selection and estimation of functional effects in the GAM~\eqref{TrueGAM} is a highly relevant problem with applications in many disciplines such as biostatistics, ecology or economics \citep{Woo2017b,FahKneLan2021}. In this framework, the additive components $f_j^\ast$ are often modeled using basis expansions such as polynomials, splines or trigonometric functions. By doing so, function selection in the GAM~\eqref{TrueGAM} can be cast into a group selection problem in a generalized linear model framework with many available options such as the group lasso \citep{YuaLin2006}, the sparse additive model \citep{RavLafLiu2009}, the Bayesian group lasso \citep{KyuGilGho2010} or the nonparametric spike-and-slab lasso \citep{BaiMorAnt2022}. 

These methods, however, follow an ``all-in-all-out'' approach for function selection \citep{GuoJaeRah2022}. That is, they do not allow  to decide whether a nonlinear effect is actually necessary to model a selected covariate or whether a linear effect is already sufficient. This is suboptimal because a linear effect is easier to interpret and reduces model complexity, thereby helping to avoid overly wiggly estimates when modeling a truly linear effect in a nonlinear fashion \citep{LouBieCar2016,GuoJaeRah2022,RosRub2023}. In recent years, several methods addressing this shortcoming have been proposed. Penalized likelihood approaches include the linear and nonlinear discoverer \citep{ZhaCheLiu2011}, generalized additive model selection \citep{ChoHas2015} and the sparse partially linear additive model \citep{LouBieCar2016}. Another option is model-based boosting \citep{HofMayRob2014} and Bayesian approaches include the methods of \cite{SchFahKne2012,HuZhaLia2015,HeWan2023} as well as non-local priors \citep{RosRub2023}. For estimation, most of these methods decompose an additive component $f_j$ into a sum of the form
    \begin{align}\label{EstimationSide}
    f_j=f_{j,lin}+f_{j,nonlin}=\widetilde{x}_j\beta_j + \sum_{l=1}^{d_j} \phi_{j,l}(x_j) u_{j,l},
\end{align}
 where $\widetilde{x}_j$ is a linear function that is standardized empirically and the $\phi_{j,l},\ l=1,\dots,d_j,$ are some basis functions that are supposed to capture the nonlinear covariate effect. Applying such a decomposition to all $f_j$, for $j=1,\ldots,p,$ the additive predictor reads as follows
 \begin{align}\label{PredictorEstimationSide}
     \eta(\bm{x})=\beta_0+\sum_{j=1}^p \beta_j \widetilde{x}_j + \sum_{j=1}^p \sum_{l=1}^{d_j } \phi_{j,l}(x_j)u_{j,l}.
 \end{align}
  For the vector of predictor evaluations $\bm{\eta}=(\eta(\bm{x}_1),\dots,\eta(\bm{x}_n))^T$ we thus obtain in matrix notation
  \begin{align}\label{AdditivePredictor}
      \bm{\eta}=\beta_0\bm{1}_n+\sum_{j=1}^p \beta_j \widetilde{\bm{x}}_j + \sum_{j=1}^p \bm{Z}_j\bm{u}_j,
 \end{align}
where $\bm{1}_n=(1,\dots,1)^T$ is a column vector of $n$ ones, $\widetilde{\bm{x}}_j=(\widetilde{x}_{1j},\ldots,\widetilde{{x}}_{nj})^T$ are the standardized covariate vectors with zero mean and unit variance and $\bm{Z}_j=(\phi_{j,l}(x_{ij}))_{i=1,\dots,n,\ l=1,\dots,d_j} \in\dsR^{n\times d_j}$ are design matrices of basis function evaluations for $j=1,\ldots,p$. To achieve sparsity, the coefficients $\beta_j$ and $\bm{u}_j$ in~\eqref{AdditivePredictor} are endowed with separate shrinkage penalties or priors. 

While the procedure seems to be  straightforward at first sight, we argue that there are two important, interrelated questions to which the existing references provide only partially satisfactory answers:

\begin{itemize}
 \item Q1: Does it suffice that the linear functions $span\{1,x_j\}$ are not contained in the span of the nonlinear basis functions  $span\{\phi_{j,1},\dots,\phi_{j,d_j}\}$ or do we need an orthogonal decomposition in~\eqref{EstimationSide} such that $span \{1,x_j\}\perp  span\{\phi_{j,1},\ldots,\phi_{j,d_j}\}$?
\item Q2: What exactly are the estimands \citep[][]{Ber2008} of $f_{j,lin}$ and $f_{j,nonlin}$ in~\eqref{EstimationSide}? Phrased differently, what exactly are the true linear $f_{j,lin}^\ast$ and nonlinear effect $f^\ast_{j,nonlin}$ of an additive component $f^\ast_{j}$ in model~\eqref{TrueGAM}? 
\end{itemize}

 Regarding Q1: Some of the references do impose orthogonality in~\eqref{EstimationSide}, others do not. \cite{SchFahKne2012} and \cite{RosRub2023}, for instance, explicitly enforce empirical orthogonality in~\eqref{EstimationSide} arguing that this leads to a better effect separation or gains in power, respectively. \cite{HofMayRob2014} and \cite{GuoJaeRah2022}, in contrast, use mixed model representations (MMRs) of penalized splines. \cite{GuoJaeRah2022} use the spectral decomposition of the penalty matrix as suggested by \cite{FahKneLan2004}, whereas \cite{HofMayRob2014} use Eiler's transformation \citep{Eil1999} based on the P-splines differences matrix \citep{EilMar1996}. In both cases, the resulting spline basis functions are not orthogonal to the linear functions. 
 \cite{HuZhaLia2015} use centered truncated power series functions for the nonlinear basis functions. These functions are not orthogonal either but in general even highly correlated with the linear functions. 

 Regarding Q2: Remarkably, only \cite{ZhaCheLiu2011} provide a rigorous definition of the true linear and nonlinear effect of an additive component $f_j^\ast$, while such a definition seems to be missing in the other references. 
However, \cite{ZhaCheLiu2011} use a reproducing kernel Hilbert framework as in \cite{Wah1990}, which does not easily translate to general basis expansion approaches. We opt for a different strategy based on projections, which appears more natural in this context, and provides elegant answers to both questions above. 

More specifically, we define the true linear effect of an additive component $f^\ast_j$ in~\eqref{TrueGAM} as the $\mathcal{L}^2(\dsP^{X_j})$-projection onto the linear functions. With this, the true linear effect is defined as the unique linear function that is as close as possible to $f^\ast_j$ in the sense of the $\mathcal{L}^2(\dsP^{X_j})$-norm. The true nonlinear effect is defined as the residual of this projection. 

For estimation, we expand the nonlinear effects in Demmler-Reinsch (DR) spline bases. The DR basis originates from the smoothing splines literature~\citep{DemRei1975} and is characterized by simultaneous orthogonality with respect to the empirical inner product and the differential semi-inner product associated with the smoothing spline roughness penalty \citep[see][Eq.~2.1]{Spe1985}. For the computationally much more convenient P-splines \citep{EilMar1996,LanBre2004} with less spline knots than observations, the DR basis is not as popular as the MMRs of \cite{Eil1999,FahKneLan2004}.   This is presumably because previous constructions of the DR basis in this context were either too restrictive \citep[e.g.,][]{NycCum1996} or too computationally intensive \citep[e.g.,][]{SchFahKne2012}. We suggest a new construction of the DR basis for P-splines, which is less restrictive and more efficient than the previous suggestions. 

To perform Bayesian effect selection, we endow the linear coefficients and the DR coefficient vectors with the recently proposed normal beta prime spike and slab (NBPSS) prior of \cite{KleCarKne2021}. NBPSS is a continuous group selection prior similar to the popular spike-and-slab group lasso (SSGL) prior of \cite{BaiMorAnt2022}. We compare the two priors in simulations and observe that NBPSS shows much better Markov chain Monte Carlo (MCMC) mixing for the latent binary selection indicators, especially if the group dimensions are not small. To provide an explanation, we establish new theoretical results for the two priors. In particular, we derive and investigate the implied spike and slab of the Euclidean norm of the group coefficient vectors. We find that the overlap is much bigger for NBPSS, providing a compelling explanation for its superior mixing performance. 

 Combining the DR bases and NBPSS, we obtain a versatile fully Bayesian approach that allows to perform data-driven effect selection with uncertainty quantification in GAMs and beyond. We illustrate the efficacy  of the developed methodology along the geoadditive Cox model \citep{HenBreFah2006} using the leukemia survival data of \cite{HenShiGor2002}. In addition to continuous covariates, the model contains a spatial effect and we also allow for flexible Bayesian estimation of the baseline hazard rate. We conduct simulations comparing our approach with the boosting approach of \cite{HofMayRob2014} and find that we outperform these authors in terms of both, estimation and selection accuracy. 
 
In summary, the main contributions of this paper are as follows.
\begin{itemize}
    \item We provide a rigorous definition of a GAM with effect decomposition based on projections and highlight the importance of orthogonality to achieve consistent effect estimation.
    \item We introduce a new construction of the DR basis for P-splines. This construction is less restrictive and more efficient  than previous suggestions. 
    \item We establish new theoretical results for the NBPSS prior. In particular, we derive the implied spike and slab of the Euclidean norm, which provides a compelling explanation for its superior MCMC mixing compared to the SSGL prior for large dimensional groups.
    \item We illustrate the developed methodology in the geoadditive Cox model. In addition to linear and nonlinear effects of continuous covariates, the model contains several dummy variables and a spatial effect, all of which are subject to data-driven effect selection.  
\end{itemize}

The rest of this paper is structured as follows.
In Section~\ref{sec:GAMED} we formally introduce the GAM with effect decomposition based on projections and detail our new construction of the DR basis for P-splines. Moreover, we demonstrate that the DR basis is well-suited for the GAM with effect decomposition, whereas the MMR yields heavily biased effect estimates. In Section~\ref{sec:BESEL} we review the NBPSS prior for Bayesian effect selection and  establish new theoretical results for the prior in Section~\ref{sec:TheoreticalRestultsNBPSS}. Section~\ref{sec:TTE}  illustrates the developed methodology in the context of survival data, while Section~\ref{sec:Discussion} concludes with a discussion. The supplementary material contains several appendices with further technical details, simulation results and proofs of our theoretical results.

\section{GAM with Effect Decomposition Based on Projections}\label{sec:GAMED}
 The cornerstone of our method is to  decompose a true additive component $f^\ast_j$ in~\eqref{TrueGAM} into the sum $f^\ast_j=f^\ast_{j,lin}+f^\ast_{j,nonlin}$, where $f^\ast_{j,lin}$ is the $\mathcal{L}^2(\dsP^{X_j})$-projection of $f^\ast_j$ onto the linear functions $\mathit{span}\{1,x_j\}$ and $f^\ast_{j,nonlin}$ is the corresponding residual. This is made precise in Definition~\ref{Def1}.
\begin{definition}[Effect decomposition]\label{Def1}
Let $X_j\in\mathcal{X}_j\subseteq\dsR$ with distribution $\dsP^{X_j}$ such that $\dsE[X_j^2]<\infty$ and $\text{Var}(X_j)>0$. Then, for a square-integrable additive component $f^\ast_j:\mathcal{X}_j\to \dsR$ we define
\begin{align}\label{EffectDecomposition}
    f^\ast_{j,lin}:=\underset{g_j\in \mathit{span}\{1,x_j\}}{\text{argmin}}\ \dsE[(f^\ast_j(X_j)-g_j(X_j))^2] \quad \text{and} \quad f^\ast_{j,nonlin}:=f^\ast_j-f^\ast_{j,lin}.
\end{align}
We refer to $f_{j,lin}^\ast$ as true linear effect and to $f_{j,nonlin}^\ast$ as true nonlinear effect. Since $f_j^\ast=f^\ast_{j,lin}+f^\ast_{j,nonlin}$, we also refer to $f_j^\ast$ as true overall effect.
\end{definition}
\begin{figure}[htbp]
        \centering
        \includegraphics[width=\textwidth]{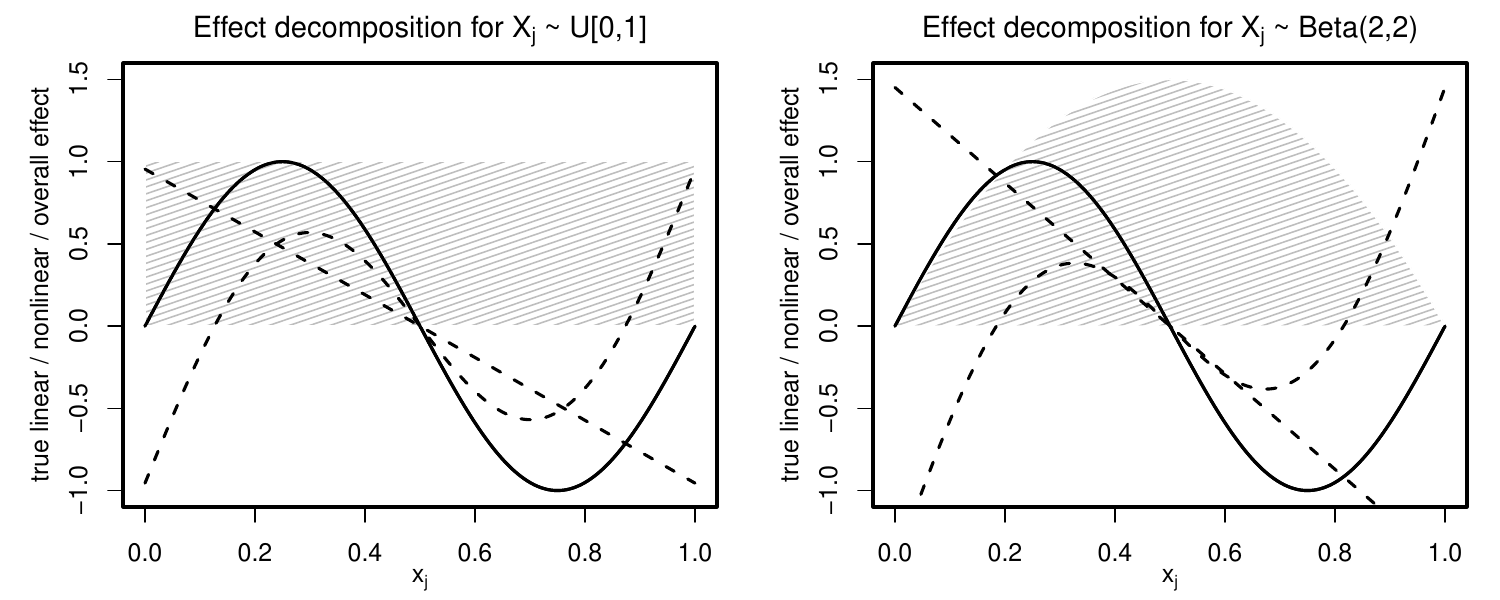}
        \caption{Effect decomposition based on $\mathcal{L}^2(\dsP^{X_j})$-projections. Shown are the true additive component $f^\ast_j=\sin(2\pi x_j)$ (solid) together with its true linear and nonlinear effects (dashed) under a uniform design $X_j\sim U[0,1]$ (left) and a beta design $X_j\sim \mathit{Beta}(2,2)$ (right). The shaded gray area depicts the corresponding design density.}
        \label{fig:ExampleEffectDecomposition}
    \end{figure}

For illustration, consider the function $f^\ast_j=\sin (2\pi x_j)$. Under a uniform design $X_j\sim U[0,1],$ we obtain the decomposition
    \begin{align*}
        f^\ast_j= f^\ast_{j,lin}+f^\ast_{j,nonlin}=\{-6/\pi(x_j-1/2)\}+\{\sin (2\pi x_j)+6/\pi(x_j-1/2)\}.
    \end{align*}
    Under a beta design $X_j\sim \mathit{Beta}(2,2),$ we obtain in contrast
    \begin{align*}
        f^\ast_j= f^\ast_{j,lin}+f^\ast_{j,nonlin}=\{-90/\pi^3 (x_j-1/2)\}+\{\sin (2\pi x_j)+90/\pi^3 (x_j-1/2)\}.
    \end{align*}
The difference is because the $\mathit{Beta}(2,2)$ distribution puts more weight into the center of the unit interval, which can best be understood graphically, see Figure~\ref{fig:ExampleEffectDecomposition}.

\begin{definition}[GAM with effect decomposition]
Applying decomposition~\eqref{EffectDecomposition} to all additive components of a GAM, we obtain a \textit{GAM with effect decomposition}, where the true additive predictor reads as follows
\begin{align}\label{GAMwithED}
\eta^\ast(\bm{X})=\beta^\ast_0 +\sum_{j=1}^p \beta^\ast_j\widetilde{X}_j+\sum_{j=1}^p f^\ast_{j,nonlin}(X_j).
\end{align}
Thereby, $\widetilde{X}_j=(X_j-\dsE [X_j])/\sqrt{\text{Var}(X_j)}$ is the standardized linear function and the true nonlinear effects $f^\ast_{j,nonlin}$ are subject to $\mathcal{L}^2(\dsP^{X_j})$-orthogonality constraints of the form
\begin{align}\label{TheoreticalOrthogonality}
\dsE[f^\ast_{j,nonlin}(X_j)]=0\quad \text{and}\quad\dsE[f^\ast_{j,nonlin}(X_j)X_j]=0,\ j=1,\dots,p.
\end{align}
\end{definition}

\begin{remark}
The natural counterpart of the constraints~\eqref{TheoreticalOrthogonality} is empirical orthogonality on the estimation side. Thus, the effect decomposition~\eqref{EffectDecomposition} has implicitly been used by several authors including \cite{SchFahKne2012,ChoHas2015,RosRub2023}, but this has not been made precise.
However, having a precise definition of the true linear and nonlinear effect is important for three reasons.
 First, having a precise definition allows us to establish a consistency result of the form $\widehat{f}_{j,lin}\overset{\dsP}{\to} f_{j,lin}^\ast$ and $\widehat{f}_{j,nonlin}\overset{\dsP}{\to} f_{j,nonlin}^\ast$ in Theorem~\ref{Prop:Consistency}. To this end, we exploit that the suggested decomposition~\eqref{EffectDecomposition} is closely related to the functional ANOVA decomposition of~\cite{Sto1994} and \cite{Hua1998b}, where a multidimensional function is decomposed into main effects and interactions. This connection allows us to use high-level results from \cite{Hua1998b} to prove our consistency result. 
   Second, we can compute effectwise root mean squared errors (RMSEs) and missclassification rates in simulation settings. For instance, if we supply the function $f^\ast_j=\sin(2\pi x_j)$ under a uniform design, we can compare the estimated linear effect $\widehat{f}_{j,lin}$ with its true counterpart $f^\ast_{j,lin}=-6/\pi(x_j-1/2)$ and similarly for the nonlinear effect. 
    Consequently, and third, separate interpretation of $\widehat{f}_{j,lin}$ and $\widehat{f}_{j,nonlin}$ becomes possible. 
\end{remark}

For estimation, we expand the nonlinear effects in a new construction of the Demmler-Reinsch (DR) basis for P-splines. We detail this in the following together with a proof that the DR basis allows for consistent effect estimation in the GAM with effect decomposition~\eqref{GAMwithED}.

\subsection{A new construction of the DR basis for P-splines}\label{sec:NewConstructionDRB}
Let $X_j$ be a continuous covariate with realized vector $\bm{x}_j=(x_{1j},\dots,x_{nj})^T$.
The construction of the design matrix $\bm{Z}_j\in\dsR^{n\times d_j}$ is based on the following three steps:
\RestyleAlgo{ruled}
\begin{algorithm}
\caption{New DR basis for P-splines}\label{alg:three}
\SetKw{KwOne}{1. Set up the P-spline design:}
\SetKw{KwTwo}{2. Compute the transition matrix:}
\SetKw{KwThree}{3. Perform the change-of-basis:}
\vspace{0.5em}
\KwOne{} Set up the B-spline design matrix $\bm{B}_j\in\dsR^{n\times (d_j+2)}$ and the associated roughness penalty matrix $\bm{K}_j\in\dsR^{(d_j+2)\times (d_j+2)}$. By default, we use cubic B-splines ($m=3$) with equidistant knots in the range of $\bm{x}_j$ and a second order differences penalty matrix $\bm{K}_j=\bm{\Delta}_j^T\bm{\Delta}_j$ (see Appendix A for details).

\KwTwo{} Set up the constraint matrix $\bm{C}_j=(\bm{1}_n,\bm{x}_j)^T\bm{B}_j\in\dsR^{2\times (d_j+2)}$ and compute a singular value decomposition (SVD) of $\bm{C}_j$ to obtain a basis $\bm{V}_{0,j}$ of the nullspace $ker(\bm{C}_j)$. 
    Compute $\widetilde{\bm{K}}_j=\bm{V}_{0,j}^T\bm{K}_j\bm{V}_{0,j}$, $\widetilde{\bm{G}}_j=\bm{V}_{0,j}^T(\bm{B}_j^T\bm{B}_j/n)\bm{V}_{0,j}\in\dsR^{d_j\times d_j}$   and solve the generalized eigenvalue problem \citep[][Chapter 15]{Par1998} for the pair of matrices $(\widetilde{\bm{G}}_j,\widetilde{\bm{K}}_j)$. Define $\bm{A}_j$ as the matrix of columnwise generalized eigenvectors and the transition matrix $\bm{T}_j=\bm{V}_{0,j}\bm{A}_j$.

\KwThree{} Set $\bm{Z}_j=\bm{B}_j\bm{T}_j$. Optionally rescale $\bm{Z}_j$ by a positive scalar, such that $trace(\bm{Z}_j^T\bm{Z}_j/n)=1$ and adjust $\bm{T}_j$ accordingly.
\vspace{1em}
\end{algorithm}

\begin{figure}[H]
    \centering
  \includegraphics[width=\textwidth]{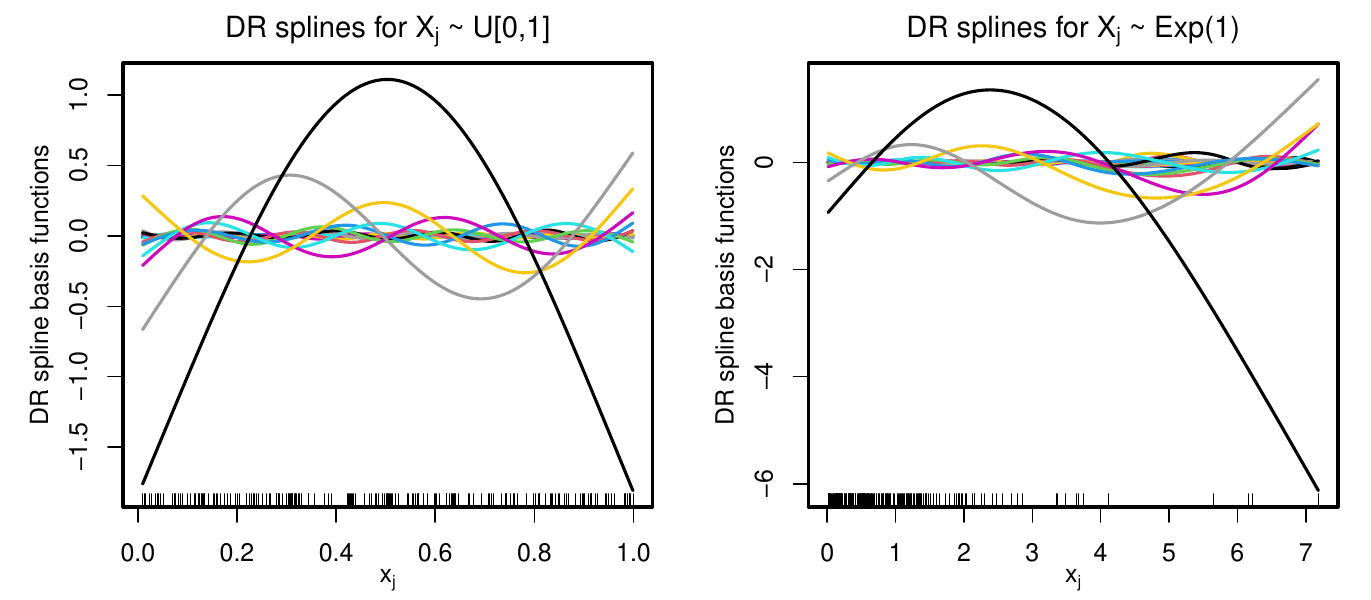}
    \caption{Illustration DR basis. Shown are $d_j=25$ DR spline basis functions for $n=200$ uniformly (left) and exponentially distributed design points $x_{ij}$ (right).}
    \label{fig:plotDRB}
\end{figure}

The resulting basis functions are empirically orthogonal in the sense that $\bm{Z}_j^T\bm{Z}_j$ is diagonal, and the corresponding roughness penalty matrix is a scalar multiple of the identity matrix $\bm{I}_{d_j}$. In addition, $(\bm{1}_n,\bm{x}_j)^T\bm{Z}_j=\bm{0}$, which will be shown to be essential for consistent effect estimation in the subsequent Section~\ref{sec:Consistency}. Figure~\ref{fig:plotDRB} illustrates the resulting basis functions for a uniform and an exponential covariate design. 

\begin{remark}
The present construction of the DR basis has three major advantages compared to existing suggestions in the literature:
\begin{enumerate}
\item It is less restrictive than the approaches of \cite{NycCum1996,Rup2002,ClaKriOps2009,Woo2017b,HeWan2023}, which 
all require $\bm{B}_j$ to have full rank. In contrast, $\bm{B}_j$ can be rank-deficient for the present construction. This is important as $\bm{B}_j$ is occasionally rank-deficient in applications when using P-splines with equidistant knots. Following a classical result by \cite{SchWhi1953}, the spline knots and the covariate values $x_{ij},\ i=1,\dots,n,$ need to satisfy an interlacing condition for $\bm{B}_j$ to have full rank, which is not always satisfied in practice \citep[see also][Chapter 14]{Boo2001}. We only require that $(\bm{1}_n,\bm{x}_j)$ has full rank, which is already the case if there exist two distinct covariate values.
\item The present construction is much more efficient than the approach of \cite{SchFahKne2012}, which relies on a truncated SVD of the implied $n\times n$  prior covariance matrix $\bm{B}_j \bm{K}_j^{+} \bm{B}_j^T$, where  $\bm{K}_j^+$ denotes the Moore-Penrose pseudoinverse of $\bm{K}_j$. This is because we only need to factorize matrices of sizes $2\times (d_j+2)$ and  $d_j\times d_j$, whereas \cite{SchFahKne2012} need to handle a matrix of size $n\times n$, and usually $d_j\ll n$ for P-splines \citep{EilMar1996}. In addition, the transition matrix $\bm{T}_j$ is directly available for the present approach facilitating predictions at test points, which is rather cumbersome for \cite{SchFahKne2012}.  
\item Algorithm \ref{alg:three} is general in the sense that it can also be applied for P-splines of a different spline degree ($m\geq 1$) or other types of penalized splines such as O'Sullivan penalized splines with integrated squared second derivative as roughness penalty \citep{OS1986,WanOrm2008}. In addition, the underlying ideas can also be used to construct suitable design matrices for spatial effects (for both discrete and continuous spatial data; see Appendix C.2 for details).
\end{enumerate}
\end{remark}

\subsection{Consistent effect estimation using the DR basis}\label{sec:Consistency}
\cite{Sto1986} establishes a general consistency result for the maximum likelihood additive spline estimator in the GAM~\eqref{TrueGAM}. It is shown that if certain regularity conditions are satisfied, then the estimated additive components $\widehat{f}_j$ converge to their true counterparts $f_j^\ast$, that is,~$\|\widehat{f}_j-f_j^\ast\|_{j}\overset{\dsP}{\to} 0$ as $n\to \infty$, where $\|\cdot\|_j$ denotes the $\mathcal{L}^2(\dsP^{X_j})$-norm. To support the suitability of the DR basis for estimation in the GAM with effect decomposition~\eqref{GAMwithED}, we next extend the result of \cite{Sto1986} to the present situation.

\begin{theorem}[Consistency] \label{Prop:Consistency}
Assume that the regularity conditions of Theorem 2 in \cite{Sto1986} hold with spline degree $m\geq 1$. Let the additive components used for maximization of the likelihood have the form $f_j=f_{j,lin}+f_{j,nonlin}=\widetilde{x}_j\beta_j+\sum_{l=1}^{d_j}\phi_{j,l}u_{j,l}$ with DR basis functions $\phi_{j,l},\ l=1,\dots,d_j,$ for the nonlinear effects. Then it holds:
\begin{align*}
    \|\widehat{f}_{j,lin}-f^\ast_{j,lin}\|_j \overset{\dsP}{\to} 0 \quad \quad\text{and}\quad\quad  \|\widehat{f}_{j,nonlin}-f^\ast_{j,nonlin}\|_j\overset{\dsP}{\to} 0,
\end{align*}
for $j=1,\ldots,p,$ as $n\to \infty$.
\end{theorem}

A proof of Theorem~\ref{Prop:Consistency} is provided in Appendix B. 

\begin{remark} Theorem~\ref{Prop:Consistency} shows that consistent estimation of the true linear and nonlinear effects in the GAM with effect decomposition~\eqref{GAMwithED} is feasible when using the DR basis from Algorithm~\ref{alg:three} to model the nonlinear effects. Owing to the generality of the result of \cite{Sto1986}, Theorem~\ref{Prop:Consistency} holds for several conditional response families $Y\mid \bm{X}$ including a Gaussian, logistic or Poisson additive model. The proof of Theorem~\ref{Prop:Consistency} is not trivial as the effect decomposition of $\widehat{f}_j$ is based on the empirical covariate distribution $\dsP_n^{X_j}=1/n\sum_{i=1}^n \delta_{X_{ij}}$, while the decomposition of $f_j^\ast$ is based on its theoretical counterpart $\dsP^{X_j}$. This leads to similar challenges as in the study of empirical processes \citep{Gee2000}, and we use a general result from \cite{Hua1998b} to show that the empirical semi-inner product and its theoretical counterpart are uniformly close on the approximating sieve of spline spaces (see Appendix B for details).
\end{remark}

\subsection{Comparison of DR basis and MMR}
To demonstrate the practical relevance of Theorem~\ref{Prop:Consistency}, we next compare the DR basis and the commonly used MMR of \cite{FahKneLan2004} in an illustrative example (see Appendix C.1 for details on the MMR and the closely related Eiler's transformation). For conciseness, we limit the presentation to the most relevant aspects, further details on the comparison are provided in Appendix C.3. Consider the model
 \begin{align*}
        Y_i=f^\ast(X_i)+\epsilon_i,\ \epsilon_i\overset{iid}{\sim} N_1(0,\sigma^2),
    \end{align*}
     with $X_i\overset{iid}{\sim} U[0,1]$ and $f^\ast(x)=\sin(2\pi x)+(x-1/2),$ as well as $\sigma^2=1$ and $n=500$.
  For estimation, we use a predictor of the form~\eqref{AdditivePredictor} with $p=1$, where we use either the DR basis or the MMR for the design matrix $\bm{Z}$ of the nonlinear effect. 
   \begin{figure}[htbp]
    \centering
    \includegraphics[width=1.0\textwidth]{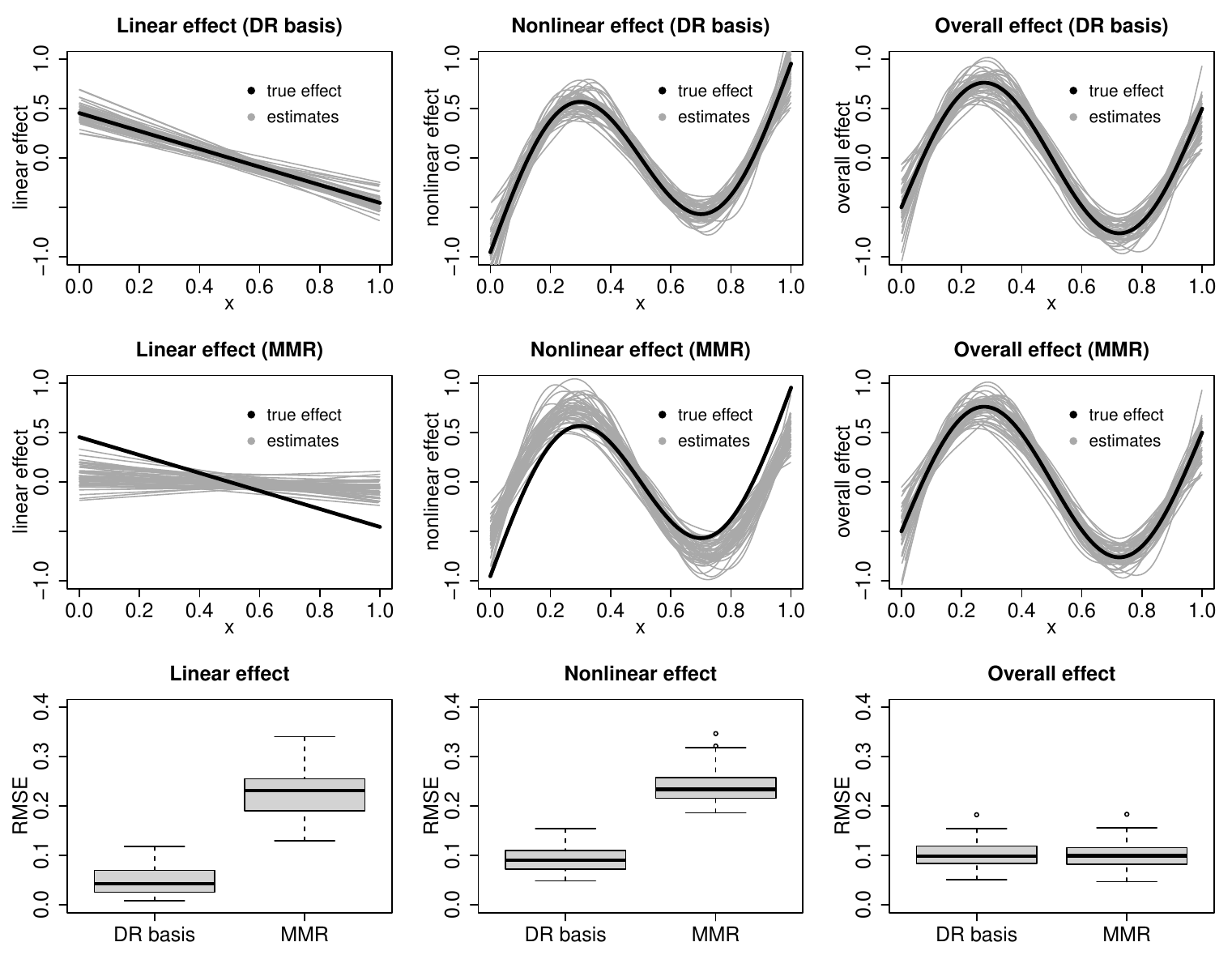}
    \caption{Comparison of DR basis and MMR. Shown are the linear/nonlinear/overall effect estimates (first/second/third column),  across the $R=50$ replicate data sets when using the DR basis (first row) or the MMR (second row) to model the nonlinear effect. The solid black lines show the true linear, nonlinear and overall effect according to Definition~\ref{Def1}. The third row shows the corresponding effectwise RMSEs.}
    \label{fig:DRBvsMMR}
\end{figure}

 Figure~\ref{fig:DRBvsMMR} depicts the resulting estimates across $R=50$ replicates. We see that the DR basis and the MMR  perform equally well for estimation of the overall effect $f^\ast$.  However, the MMR does not perform well in estimating the true linear and nonlinear effects separately. In fact, both effect estimates are highly biased for the MMR. The reason is that $(\bm{1}_n,\bm{x})^T\bm{Z}=\bm{0}$ for the DR basis but not for the MMR. In Appendix~C.3 we conduct several additional analyses showing inter alia that the bias for the MMR persists as $n$ increases and that Eiler's transformation (the default in the popular R package \texttt{mboost}) leads to similar problems. In general, we conclude that empirical orthogonality, i.e. $(\bm{1}_n,\bm{x}_j)^T\bm{Z}_j=\bm{0}$, is crucial to achieve satisfactory estimation performance in the GAM with effect decomposition~\eqref{GAMwithED}.

\section{Bayesian Effect Selection}\label{sec:BESEL}
In this section we combine the DR basis with a Bayesian group selection prior to perform data-driven effect selection. We consider a GAM with effect decomposition~\eqref{GAMwithED} and a potentially large number of continuous covariates $p$. We assume \textit{effect sparsity}, that is,~we assume that only some of the covariates have a linear effect ($\beta^\ast_j\neq 0$) and only some of the covariates have a nonlinear effect ($f^\ast_{j,nonlin}\neq 0$). Some of the covariates may have no effect on the response at all ($\beta^\ast_j=0,f^\ast_{j,nonlin}=0$). 

Our goal is to perform effect selection, that is, we aim to develop a framework that allows to detect automatically (i.e.~in a data driven manner) whether a linear or a nonlinear effect of a continuous covariate are present or not. To facilitate the exposition, we first introduce a new notation for the vector of predictor evaluations~\eqref{AdditivePredictor}. In the following we write 
\begin{align}\label{NewNotationEta}
            \bm{\eta}= \beta_0 \bm{1}_n + \sum_{j=1}^{J} \bm{\psi}_j\bm{\beta}_j=\beta_0\bm{1}_n+\bm{\psi}\bm{\beta},
\end{align}
where the $\bm{\psi}_j$ are generic design matrices of sizes $n\times d_j$ and the $\bm{\beta}_j\in\dsR^{d_j}$ are coefficient vectors to be estimated. Thereby, $\bm{\psi}_j=\widetilde{\bm{x}}_j$ for a linear effect and $\bm{\psi}_j=\bm{Z}_j$ from Algorithm~\ref{alg:three} for a nonlinear effect. Moreover, $\bm{\psi}=(\bm{\psi}_1,\dots,\bm{\psi}_J)$ and $\bm{\beta}=(\bm{\beta}_1^T,\dots,\bm{\beta}_J^T)^T$. In terms of~\eqref{NewNotationEta}, effect sparsity simply corresponds to group sparsity, that is, we assume that $\bm{\beta_j}=\bm{0}$ for some of the $j\in\{1,\dots,J\}$. Many different approaches such as penalized likelihood or boosting can be used to tackle the resulting group selection problem. In this paper, we opt for a fully Bayesian approach with MCMC sampling. To this end, we endow the $\bm{\beta}_j$ with the recently proposed normal beta prime spike and slab (NBPSS) prior of \cite{KleCarKne2021}. 

\subsection{Normal beta prime spike and slab prior}\label{sec:NBPSS}
We introduce positive variance parameters $\tau_j^2$ and binary selection indicators $\gamma_j\in\{0,1\}$ as well as effect-specific weights $\omega_j\in (0,1)$. Following \cite{KleCarKne2021} we use a prior of the form
\begin{equation}\begin{aligned}\label{NBPSSPrior}
    \bm{\beta}_j\mid \tau_j^2&\sim N_{d_j}(\bm{0},\tau_j^2\bm{I}_{d_j})\\
    \tau_j^2\mid \gamma_j&\sim \mathit{SBP}(a_j,b_j,c_{j,\gamma_j}),\\
    \gamma_j\mid \omega_j &\sim \mathit{Bernoulli}(\omega_j),\\
    \omega_j &\sim \mathit{Beta}(a_{\omega_j},b_{\omega_j}), 
\end{aligned}\end{equation}
for $j=1,\dots,J,$ with  scale parameters $c_{j,0}\ll c_{j,1}$ for the spike and the slab, respectively. Above, $\mathit{SBP}(a,b,c)$ denotes the scaled beta prime distribution with shape parameters $a$ and $b$ and scale parameter $c$ whose density function is 
\begin{align*}
    \mathit{SBP}(x;a,b,c)=1/({c\ B(a,b)})\ {(x/c)^{(a-1)}(1+x/c)^{-(a+b)}},\ x>0,
\end{align*}
where $B(a,b)$ is the beta function \citep[see][for details]{PerPerRam2017}. To complete the prior specification, we place an improper uniform prior $p(\beta_0)\propto 1$ on the intercept and a suitable prior on the dispersion parameter if present (e.g.,  Jeffreys' prior $p(\sigma^2)\propto 1/\sigma^2$ in the Gaussian model).

\subsection{Prior hyperparameters}
We follow \cite{KleCarKne2021} and use $a_j=1/2$ and $b_j=5$ for the shape parameters of the scaled beta prime distribution. To elicit suitable values for the scale parameters $c_{j,0}$ and $c_{j,1}$, we use a prior scaling approach. The key idea is to choose $c_{j,0}$ and $c_{j,1}$ such that prior effect draws $\bm{f}_j=\bm{\psi}_j\bm{\beta}_j$ from both, the spike and the slab have a reasonable size (see Appendix~E.2 for details). For the $\omega_j$ we use uniform priors with $a_{\omega_j}=b_{\omega_j}=1$ by default. Our implementation also supports a global $\omega$, which may allow for better adaption to the unknown sparsity level in high-dimensional settings \citep{Roc2018,BaiMorAnt2022}.

\subsection{Posterior inference}
Effect selection and estimation is based on the posterior distribution, which by Bayes' rule, has density proportional to
\begin{align*}
    p(\beta_0,\bm{\beta},\bm{\tau}^2,\bm{\gamma},\bm{\omega}\mid \mathcal{D})\propto p(\mathcal{D}\mid \beta_0,\bm{\beta}) \prod_{j=1}^J p(\bm{\beta}_j\mid \tau_j^2) p(\tau_j^2\mid \gamma_j) p(\gamma_j\mid \omega_j) p(\omega_j),
\end{align*}
where $\bm{\tau}^2=(\tau_1^2,\ldots,\tau_J^2)^T$, $\bm{\gamma}=(\gamma_1,\ldots,\gamma_J)^T$ and $\bm{\omega}=(\omega_1,\ldots,\omega_J)^T$. 
As the posterior is analytically intractable, we use MCMC to sample from it. A schematic overview of our sampler is shown in Algorithm~\ref{alg:two}, further details are provided in Appendix~E.3.

\RestyleAlgo{ruled}
\begin{algorithm}
\caption{Schematic MCMC sampler for NBPSS}\label{alg:two}
\vspace{1em}
For $t=1,\dots,T$:
\begin{itemize}
    \item Sample $\bm{\beta}^{(t)}$ using a Metropolis-Hastings step with IWLS proposal \citep{Gam1997}.
    \vspace{-0.1em}
    \item For $j=1,\dots,J$: Sample $(\tau_j^2)^{(t)}$ using a slice step \citep{Nea2003}.
    \vspace{-0.1em}
    \item For $j=1,\dots,J$: Sample $\gamma_j^{(t)}$ using a Gibbs step.
    \vspace{-0.1em}
    \item  For $j=1,\dots,J$: Sample $\omega_j^{(t)}$ using a Gibbs step.
\end{itemize}
\end{algorithm}

\subsection{Effect selection and estimation}
We use the marginal posterior inclusion probabilities $\dsP(\gamma_j=1\mid \mathcal{D})$ for effect selection. Specifically, we say that the $j$-th effect $\bm{f}_j=\bm{\psi}_j\bm{\beta}_j$ is present if $\dsP(\gamma_j=1\mid \mathcal{D}) \geq 0.5,$
which corresponds to the median probability model \citep[][]{BarBer2004,BarBerGeo2021}. For effect estimation, we use the marginal posterior mean
        $\widehat{\bm{f}}_j=\bm{\psi_j} \widehat{\bm{\beta}}_j$ with $\widehat{\bm{\beta}}_j=\dsE[\bm{\beta}_j\mid \mathcal{D}]$,
and for uncertainty quantification we use $95\%$ posterior credible intervals (pointwise, equal-tailed).

\subsection{MCMC mixing of the binary selection indicators}
It is well-known that MCMC mixing of the binary selection indicators $\gamma_j$ is notoriously difficult for Bayesian group selection approaches, especially if the group dimensions $d_j=\mbox{dim}(\bm{\beta}_j)$ are not small \citep[cf.,][]{SchFahKne2012, KleCarKne2021,WieKneWag2021b}. In the worst case, the MCMC sampler can get stuck in the spike $\gamma_j=0$ or the slab $\gamma_j=1$, leading to unreliable MCMC inference for the posterior inclusion probabilities $\dsP(\gamma_j=1\mid \mathcal{D})$. Following \cite{SchFahKne2012}, we approximate the $\dsP(\gamma_j=1\mid \mathcal{D})$ using the average of the conditional posterior inclusion probabilities $1/T\ \sum_{t=1}^T \dsP(\gamma_j^{(t)}=1\mid \cdot)$ instead of the average of the indicators $1/T \sum_{t=1}^T \gamma_j^{(t)}$. This corresponds to Rao-Blackwellization, which is usually slightly more efficient \citep{RobRob2021}. We investigate the accuracy of the Rao-Blackwellized MCMC approximation in Appendix~E.4. In line with the findings of~\cite{KleCarKne2021}, we observe that MCMC mixing is very satisfactory for NBPSS, even for large dimensional groups. For comparison, we also investigate the mixing when using the popular SSGL prior of \cite{BaiMorAnt2022} as a potential alternative to the NBPSS prior, but we find that NBPSS generally yields much better MCMC mixing. In the following Section~\ref{sec:TheoreticalRestultsNBPSS} we establish new theoretical results, which may explain the favorable MCMC mixing of NBPSS compared to the SSGL.

\section{Theoretical Results}\label{sec:TheoreticalRestultsNBPSS}
In this section we derive the following new theoretical results for the NBPSS prior.
\begin{enumerate}
    \item A closed form representation for the marginal spike $p(\bm{\beta}_j\mid \gamma_j=0)$ and slab $p(\bm{\beta}_j\mid \gamma_j=1)$ of the group coefficient vector $\bm{\beta}_j\in \dsR^{d_j}$ with variance parameter $\tau_j^2$ integrated out.
    \item The implied spike $p(r_j\mid \gamma_j=0)$ and slab $p(r_j\mid \gamma_j=1)$ of the Euclidean norm $r_j=\|\bm{\beta}_j\|_2$.
\end{enumerate}
The implied spike and slab of the Euclidean norm are particularly interesting as they allow us to visualize the NBPSS prior for arbitrary group dimension $d_j\in \dsN$, while the marginal spike and slab can only be plotted for $d_j\in\{1,2\}$.

\subsection{Marginal spike and slab}
To derive the marginal spike and slab, we need to compute the integral
\begin{align*}
        p(\bm{\beta}\mid \gamma) &=\int_0^\infty N_{d}(\bm{\beta};\bm{0},\tau^2 \bm{I}_{d})\ SBP(\tau^2;a,b,c_\gamma)d\tau^2\\
        &=\dfrac{1}{(2\pi)^{d/2}c_\gamma B(a,b)}\int_0^\infty  (\tau^2)^{-d/2}\exp(-\|\bm{\beta}\|_2^2/(2\tau^2)) (\tau^2/c_\gamma)^{a-1}\ (1+\tau^2/c_\gamma)^{-(a+b)} d\tau^2,
    \end{align*}
    where we omit the group index $j\in\{1,\dots,J\}$ to simplify the notation. \cite{HerHerDup2013} refer to the density for the special case $a=b=1/2$ as \emph{group horseshoe} density and write that it is not possible to evaluate the above integral in closed form. \cite{KleCarKne2021} use numerical integration to compute and visualize $p(\bm{\beta}\mid \gamma)$ for $d\in \{1,2\}$. Proposition~\ref{prop:MarginalSpikeAndSlab} contradicts \cite{HerHerDup2013} and provides a closed-form expression for $p(\bm{\beta}\mid\gamma)$.

    \begin{proposition}[Marginal spike and slab]~\label{prop:MarginalSpikeAndSlab}
        \begin{align}\label{eq:MarginalSpikeSlab}
    p(\bm{\beta}\mid \gamma)&= \dfrac{\Gamma(b+d/2)}{(2\pi c_\gamma)^{d/2}B(a,b)}\ \mathcal{U}(b+d/2,-a+d/2+1,\|\bm{\beta}\|_2^2/(2c_\gamma)),\ \bm{\beta}\in \dsR^d,
\end{align}
where $\mathcal{U}$ is Kummer's U function \citep[see, e.g.,][Chapter 13]{AbrSte1972}.
    \end{proposition}

A proof of Proposition~\ref{prop:MarginalSpikeAndSlab} is provided in Appendix~D.1. 

\begin{figure}[htbp]
    \centering
    \includegraphics[width=\textwidth]{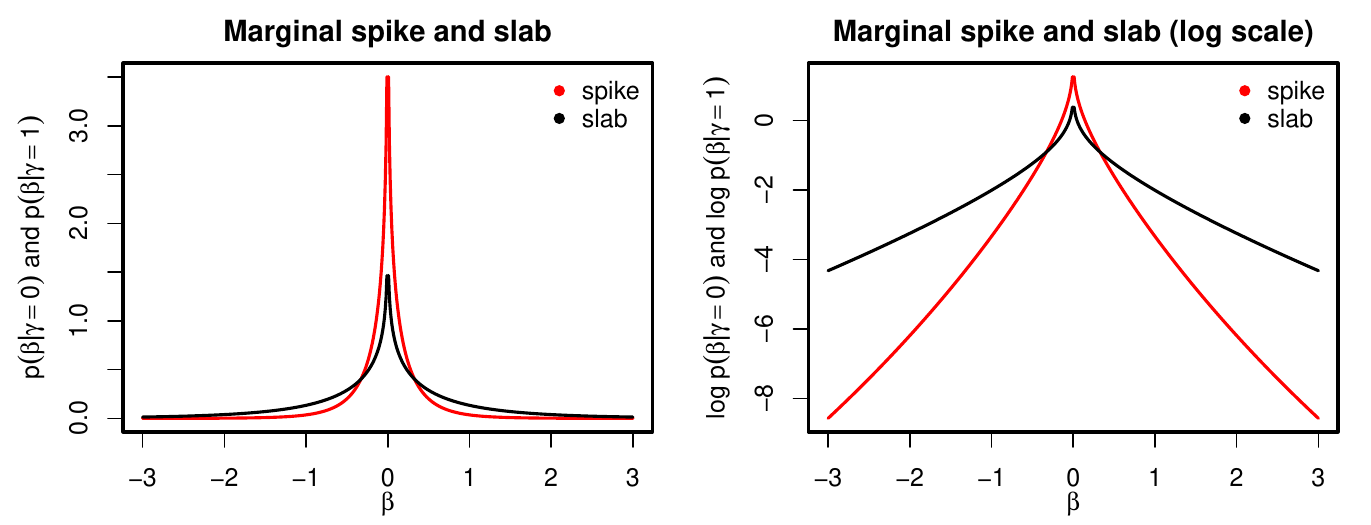}
    \caption{Theoretical results I. Shown are the marginal spike and slab~\eqref{eq:MarginalSpikeSlab} for $d=1$ on the original scale (left) and the log scale (right). To evaluate Kummer's U function we use the R package \texttt{reticulate}, allowing us to access the function \texttt{hyperu} in the Python module \texttt{scipy.special}.}
    \label{fig:enter-label}
\end{figure}

\begin{remark}
\cite{ArmDunLee2013b} state the density~\eqref{eq:MarginalSpikeSlab} for the univariate case ($d=1$). They refer to it as \emph{horseshoe-like} but do not provide a derivation. A derivation for the univariate case can be found in \cite{AguBur2023}. Proposition \ref{prop:MarginalSpikeAndSlab} generalizes this result to arbitrary dimension $d\in \dsN$ and we refer to the density~\eqref{eq:MarginalSpikeSlab} and the corresponding distribution as \emph{group horseshoe-like}. Figure \ref{fig:enter-label} shows an illustration for the univariate case $d=1$.    
\end{remark}

\subsection{Implied spike and slab of the norm}
Next, we derive the implied spike and slab of the Euclidean norm $r=\|\bm{\beta}\|_2$, which will allow us to visualize the NBPSS prior for arbitrary group dimension $d\in\dsN$. To this end, we use a general result from \cite[][Section 5]{Kel1970} for spherical distributions. With this, we obtain

\begin{proposition}[Spike and slab of the norm]~\label{Prop:SpikeSlabNorm}
\begin{align*}
    p(r\mid \gamma) =\dfrac{2\ \Gamma(b+d/2)}{(2 c_\gamma)^{d/2}\Gamma(d/2) B(a,b)}\ r^{d-1}\ \mathcal{U}(b+d/2,-a+d/2+1,r^2/(2c_\gamma)),\ r>0.
\end{align*}    
\end{proposition}

A proof of Proposition~\ref{Prop:SpikeSlabNorm} is provided in Appendix~D.2. The implied spike and slab of the norm have the following properties.

\begin{proposition}[Properties of the spike and slab of the norm]\label{prop:PropertiesImpliedSpikeAndSlab}~
    \begin{enumerate}[i)]
        \item The expected value is 
\begin{align*}
    \dsE[r\mid \gamma]= \int_0^\infty r\ p(r\mid \gamma)dr=\sqrt{2c_\gamma}\ \dfrac{B(a+1/2,b-1/2)}{B(a,b)}\dfrac{\Gamma((d+1)/2)}{\Gamma(d/2)}
\end{align*}
for $b>1/2$ and $+\infty$ for $b\leq 1/2$.
    \item The tail decay is controlled by the parameter $b$ and it holds $p(r\mid \gamma) = O(r^{-(2b+1)})$ as $r\to \infty.$
    \item The behavior at the origin is controlled by the parameter $a$ and it holds:
    \begin{align*}
        \lim_{r\to\infty} p(r\mid \gamma)= \begin{pmatrix}
            \infty ,\ a\leq 1/2  \\
            \kappa_1,\ a>1/2            
        \end{pmatrix}\quad \text{and}\quad  \lim_{r\to\infty} p(r\mid \gamma)= \begin{pmatrix}
            \infty ,\ a< 1/2  \\
            \kappa_2 ,\ a=1/2    \\        
            0,\ a>1/2
        \end{pmatrix},
    \end{align*}
    for $d=1$ and $d>1$ respectively, where $\kappa_1,\kappa_2\in(0,\infty)$ are finite positive constants.
    \end{enumerate}
\end{proposition}
A proof of Proposition~\ref{prop:PropertiesImpliedSpikeAndSlab} is provided in Appendix~D.3. 

Figure~\ref{fig:ImpliedSpikeAndSlab} shows the implied spike and slab of the Euclidean norm for our choice of shape parameters $a=1/2$ and $b=5$ for a one-dimensional group ($d=1$) and a ten-dimensional group ($d=10$). The scale parameters $c_0$ and $c_1$ were chosen such that $\dsE[r\mid \gamma=0]=0.1$ and $\dsE[r\mid \gamma=1]=1$. For comparison, we also show the implied spike and slab of the norm for the SSGL prior of \cite{BaiMorAnt2022}. In that case the marginal spike and slab are group lasso densities of the form $p(\bm{\beta}\mid \gamma) \propto \exp(-\lambda_\gamma \|\bm{\beta}\|_2),\ \bm{\beta}\in \dsR^d,$ and the implied spike and slab of the norm are gamma densities of the form $p(r\mid \gamma)\propto r^{d-1}\exp(-\lambda_\gamma r),\ r>0.$ For the behavior at the origin it holds $\lim_{r\to\infty} p(r\mid \gamma)=\lambda_\gamma$ for $d=1$ and $\lim_{r\to\infty} p(r\mid \gamma)=0$ for $d>1$.

\begin{figure}[hbtp]
    \centering
    \includegraphics[width=\textwidth]{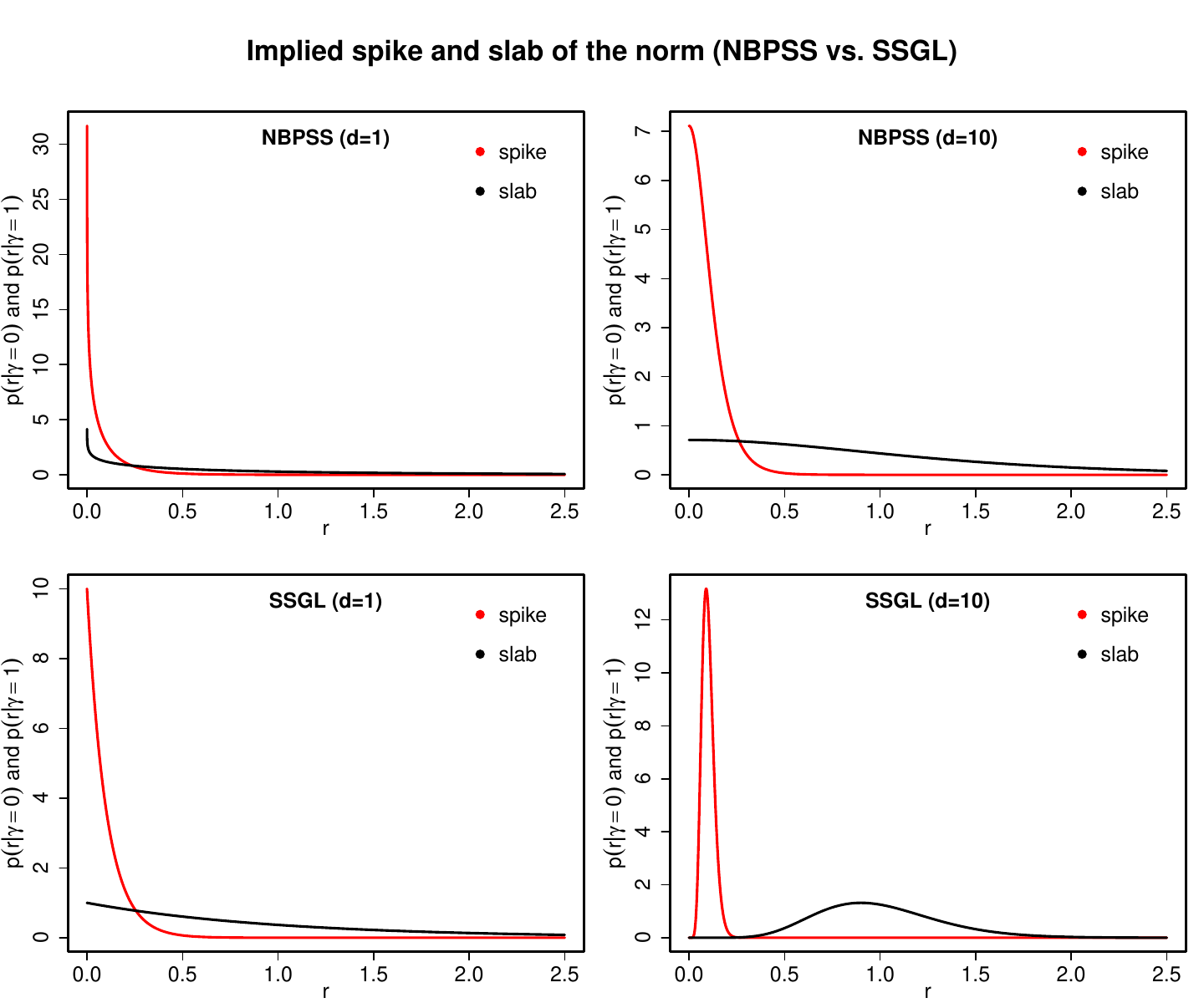}
    \caption{Theoretical results II. Shown are the implied spike and slab of the Euclidean norm for the NBPSS prior (top panel) and the SSGL prior (bottom panel). On the left we have $d=1$ and on the right we have $d=10$. The scale parameters were chosen such that $\dsE[r\mid \gamma=0]=0.1$ and $\dsE[r\mid \gamma=1]=1$ for each plot.}
    \label{fig:ImpliedSpikeAndSlab}
\end{figure}

\begin{remark}
 The implied spike and the slab of the Euclidean norm clearly overlap for NBPSS. This is true for both $d=1$ and $d=10$. For the SSGL, in contrast, the implied spike and the slab of the norm only overlap for $d=1$ but not for $d=10$. The reason is that the norm of the group lasso distribution concentrates sharply around $d/\lambda_\gamma$ as the group dimension $d$ grows. This behavior is somewhat similar to the concentration of the norm of a multivariate Gaussian distribution \citep[][Chapter 3]{Ver2018}. The group horseshoe-like distribution, in contrast, does not exhibit a similar behavior, provided that $a\leq 1/2$. To the best of our knowledge, this difference has not been highlighted in the literature before.

In Appendix~E.4 we investigate the MCMC mixing of the binary selection indicators $\gamma_j$ for both NBPSS and the SSGL and find that mixing is generally much better for NBPSS. We conjecture that the superior mixing of NBPSS can be explained by Figure~\ref{fig:ImpliedSpikeAndSlab}. This is plausible because a sufficient overlap of the spike and the slab is also necessary to achieve good MCMC mixing in the context of Bayesian variable selection \citep{GeoMcC1993,GeoMcC1997}. 
\end{remark}

\section{Application to Time-to-Event Data}\label{sec:TTE}
 In this section we illustrate the applicability of NBPSS for time-to-event data. Time-to-event-data is very common in medical studies and the Cox model \citep{Cox1972b} is one of the most popular regression models in this context \citep{GeoSeaAba2014}. Thereby, the covariates are linked to the hazard rate of the time-to-event response and in addition to categorical and continuous covariates, often also spatial information like the district or the exact place of residence of the study participants is available \citep{HenBreFah2006}. While many approaches for estimation of flexible geoadditive Cox regression models are available \citep[e.g.][]{MarAkeRue2011,Woo2017b,ZhoHanZha2020,BriElcNov2020}, only few allow for data-driven effect selection. One approach that is capable of effect selection in the geoadditive Cox model is model-based boosting \citep{HofMayRob2014}. 
 In what follows, we first extend the geoadditive Cox model of \cite{HenBreFah2006} 
to include our effect decomposition~\eqref{GAMwithED}. Then, we conduct simulations comparing the performance of NBPSS and model-based boosting before considering an application on survival with leukemia.

\subsection{The geoadditive Cox model with effect decomposition}
Suppose we have data $\mathcal{D}=\{(t_i,\delta_i,\bm{v}_i),\ i=1,\dots,n\},$ where $t_i\geq 0$ are time points, the $\delta_i\in\{0,1\}$ are binary event indicators ($\delta_i=1$ means that the event has occurred), and the $\bm{v}_i\in \dsR^p$ comprise realized covariate information of mixed type (binary, continuous, spatial) of $\mV$. Without loss of generality we assume that categorical covariates are represented through dummy variables. Similar to \cite{HenBreFah2006}, we assume that the conditional hazard rate has the form
\begin{align*}
h^\ast(t\mid \bm{V})=\exp(g^\ast_0(t)+\eta^\ast(\bm{V})).
\end{align*}
Thereby, the log baseline hazard rate $g^\ast_0(\cdot)$ is an unknown smooth function of time and $\eta^\ast(\bm{V})$ is a geoadditive predictor of the form
\begin{align}\label{TruePredictorCox}
    \eta^\ast(\bm{V})= \sum_{j=1}^{p_d}\alpha^\ast_j D_j+ \sum_{j=1}^{p_c} \beta_j^\ast \widetilde{X}_j+ \sum_{j=1}^{p_c} f^\ast_{j,nonlin}(X_j) + f^\ast_{spat}(\bm{S}),
\end{align}
with identifiability constraints $\dsE [f^\ast_{j,nonlin}(X_j)]=\dsE[f^\ast_{j,nonlin}(X_j) X_j]=0,\ j=1,\dots,p_c,$ as well as $\dsE[f^\ast_{spat}(\bm{S})]=0$. 
Some remarks are in order.
\begin{itemize}
    \item The predictor $\eta^\ast$ contains no intercept $\beta_0^\ast$ since the overall level is covered through $g_0^\ast$.
    \item The $D_j,\ j=1,\dots,p_d,$ are dummy variables with values in $\{0,1\}$. 
    \item The functional effects of the continuous covariates $X_j,\ j=1,\dots,p_c,$ are decomposed into a linear and a nonlinear effect as in~\eqref{GAMwithED}. 
    \item The rightmost term $f^\ast_{spat}(\bm{S})$ is a spatial effect for either continuous data $\bm{S}\in \Omega\subseteq \dsR^2$ or discrete regional data $\bm{S}\in \{1,\dots,R\}$. 
\end{itemize}

Crucially, we again assume effect sparsity, that is, we assume that some of the terms in~\eqref{TruePredictorCox} are negligible. Our goals are twofold: We want to identify the active effects and estimate them.

\subsection{Estimation}
For estimation, we expand the nonlinear effects and the spatial effect in DR bases (see Algorithm~\ref{alg:three} and Appendix~C.2, respectively). For the dummy variables we use the raw covariates and for the linear effects we use empirically standardized linear functions (as before). For the vector of predictor evaluations we can thus write $\bm{\eta} =\sum_{j=1}^J \bm{\psi}_j \bm{\beta}_j=\bm{\psi}\bm{\beta}$, where $J=p_d+2\times p_c+1$ is the overall number of effects. Following \cite{HenBreFah2006} we use cubic Bayesian P-splines \citep{LanBre2004} to model the log baseline hazard rate $g_0(\cdot)$. Thus, we can write $\bm{g_0}=\bm{\psi}_0\bm{\beta}_0$ for the corresponding vector of function evaluations. 
With the usual assumptions about noninformative censoring, the full likelihood is
\begin{align*}
p(\mathcal{D}\mid \bm{\beta}_0,\bm{\beta})=\prod_{i=1}^n h(t_i\mid \bm{v}_i)^{\delta_i} \exp\left(-\int_0^{t_i} h(u\mid \bm{v}_i)du\right).
\end{align*}
\noindent
To reflect our sparsity assumption, we endow the regression coefficients $\bm{\beta}=(\bm{\beta}_1^T,\dots,\bm{\beta}_J^T)^T$ with the NBPSS prior~\eqref{NBPSSPrior}. In Appendix~E.1 we derive the gradient and Hessian of the log-likelihood, which we need for efficient posterior sampling (see Appendix~E.3). As the resulting expressions contain analytically intractable integrals, we cannot evaluate them directly but need to resort to numerical approximations (see Appendix~E.1.2). 

\subsection{Numerical experiments}\label{sec:NumericalExperiments}
To investigate the performance of NBPSS in the geoadditive Cox model~\eqref{TruePredictorCox}, we compare our method with the boosting approach of \cite{HofMayRob2014} as implemented in the R package \texttt{mboost}. To the best of our knowledge, this is the only directly available approach allowing for data-driven selection of linear, nonlinear and spatial effects in the geoadditive Cox model. The following two questions are of particular interest: 1.~Does the DR basis lead to better selection and estimation performance than the MMR or Eiler's transformation? 2.~Can \texttt{mboost} also benefit from using the DR basis? To answer the second question we use the flexibility of \texttt{mboost} to create new base-learners.

\paragraph{Simulation setting}
Following~\cite{HenBreFah2006} and to resemble our real data application in the subsequent Section~\ref{sec:Application}, we consider the following data generating process:
\begin{itemize}
    \item There are two dummy covariates $D_1$ and $D_2$ (each of them Bernoulli with mean $1/2$) and eight continuous covariates $X_1,\dots,X_{8}$ (each of them uniform on $[0,1]$). In addition, there is a two-dimensional spatial covariate $\bm{S}$, which is uniform on the unit square $[0,1]^2$.
    \item The time points are generated from a Weibull distribution with shape parameter $k$ and scale parameter $\exp(-\eta^\ast(\bm{V})/k)$, that is, $T\mid \bm{V}\sim \mathit{Weibull}(k, \exp(-\eta^\ast(\bm{V})/k))$, which corresponds to the conditional log hazard rate 
$\log h^\ast(t\mid \bm{V})=g_0^\ast(t)+\eta^\ast(\bm{V})=\log k +(k-1)\log t +\eta^\ast(\bm{V}).$

\item The true geoadditive predictor $\eta^\ast$ is
\begin{align}\label{TruePredictorCoxSimu}
    \eta^\ast(\bm{V})=0\times D_1+1/2\times D_2+\sum_{j=1}^8 \beta_j^\ast \widetilde{X}_j+\sum_{j=1}^8 f^\ast_{j,nonlin}(X_j)+f^\ast_{spat}(\bm{S}),
\end{align}
where $\widetilde{X}_j=\sqrt{12}(X_j-1/2)$ and $\bm{\beta}^\ast=(\beta_1^\ast,\dots,\beta_{8}^\ast)^T=(1/\sqrt{12},-2/\sqrt{12},0,0,\beta_5^\ast,-\sqrt{3}/\pi,0,0)^T$ with $\beta_5^\ast=-2\{5+\exp(3)\}/\{3\sqrt{3}\exp(3)\}$. For the true nonlinear effects we use
\begin{itemize}
    \item $f^\ast_{1,nonlin}=f^\ast_{2,nonlin}=f^\ast_{7,nonlin}=f^\ast_{8,nonlin}=0$,
    \item $f^\ast_{3,nonlin}=4(x-1/2)^2-1/3$,
    \item $f^\ast_{4,nonlin}=f^\ast_{6,nonlin}=\sin(2\pi x)+6/\pi(x-1/2)$,
    \item $f^\ast_{5,nonlin}=2\exp(-3x)-\{2/3-2/3\exp(-3)\}-\beta_5^\ast\sqrt{12}(x-1/2)$.
\end{itemize}
It is straightforward to verify that these satisfy the required identifiability constraints~\eqref{TheoreticalOrthogonality} (see Appendix F.1 for details and plots of the functions). For the spatial effect we set $f^\ast_{spat}\equiv 0$ so that there is no true spatial effect in $\eta^\ast$. With this, the true model has selection indicators
$\bm{\gamma}^\ast=(0,1\mid 1,1,0,0,1,1,0,0\mid 0,0,1,1,1,1,0,0\mid 0)^T,$ 
where we use the same order as in~\eqref{TruePredictorCoxSimu}, i.e. we first have the two dummies, followed by the linear, nonlinear and spatial effects.
\item We generate a sample of size $n=1,000$ with Weibull shape $k=3/2$ and censor the generated time points at $t_{cens}=2$ leading to a censoring rate of about $14\%$.
\end{itemize}

\paragraph{Competitors and performance measures}
We generate $R=100$ replicate data sets and apply NBPSS and \texttt{mboost} for estimation and effect selection. For both methods we consider three basis expansions based on Eiler's transformation, the MMR and the DR basis for the nonlinear effects of the continuous covariates. Throughout, we use $d_j=9$ basis functions. Other choices such as $d_j\in\{14,19\}$ led to slightly worse performance although the differences were minor. For NBPSS we use 1,000 MCMC iterations from which we discard the first 100 as burn-in. For \texttt{mboost} we use early stopping as recommended by \cite{HofMayRob2014}. To this end, we first use 500 boosting iterations and then optimize the number of boosting iterations using 25 bootstrap samples. To judge the performance of NBPSS and \texttt{mboost} in the three different variants, we compute effectwise missclassification rates $1/R\sum_{r=1}^R|\widehat{\gamma}^{(r)}_j-\gamma_j^\ast|,\ j=1,\dots,J$, as well as the overall RMSE $\|\widehat{\bm{\eta}}-\bm{\eta}^\ast\|_2/\sqrt{n}$ at the observed covariates. 
\paragraph{Results}
 
\begin{figure}[htbp]
    \centering
    \includegraphics[width=\textwidth]{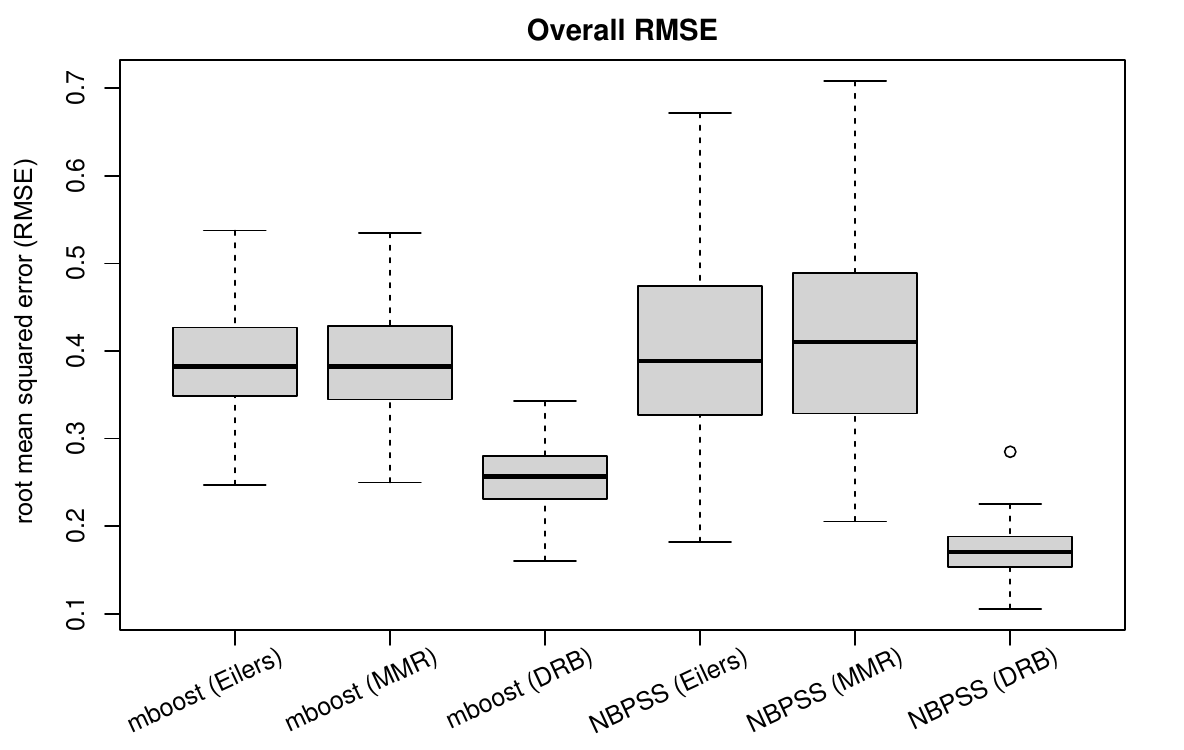}
    \caption{RMSEs. Shown is the overall RMSE $\|\widehat{\bm{\eta}}-\bm{\eta}^\ast\|_2/\sqrt{n}$ for the different variants of \texttt{mboost} and NBPSS across the $R=100$ replicates. }
    \label{fig:RMSE}
\end{figure}

\hspace{-1em}
\begin{table}[htbp]
\centering
\setlength{\tabcolsep}{2pt}
\renewcommand{\arraystretch}{0.8}
\begin{tabular}{r|rrrrrrrrrrrrrrrrrrr|r}
 Effect & D1 & D2 & L1 & L2 & L3 & L4 & L5 & L6 & L7 & L8 & N1 & N2 & N3 & N4 & N5 & N6 & N7 & N8 & S & Av.\\ 
  \hline
  Active &0 & 1 & 1 & 1 & 0 & 0 &1 &1 &0 &0 & 0&0&1&1&1&1&0&0& 0 & 9/19\\
\hline
mboost (Eilers) & 10 & 0 & 0 & 0 & 0 & 98 & 0 & 79 & 4 & 6 & 97 & 97 & 0 & 0 & 0 & 0 & 83 & 78 & 99 & 34 \\ 
  mboost (MMR) & 11 & 0 & 0 & 0 & 0 & 97 & 0 & 80 & 5 & 5 & 97 & 97 & 0 & 0 & 0 & 0 & 84 & 75 & 97 & 34 \\ 
 mboost (DRB) & 6 & 0 & 0 & 0 & 32 & 27 & 0 & 0 & 14 & 28 & 92 & 91 & 0 & 0 & 0 & 0 & 89 & 86 & 96 & 30 \\ 
 NBPSS (Eilers) & 2 & 0 & 0 & 0 & 3 & 100 & 0 & 84 & 2 & 1 & 4 & 6 & 0 & 0 & 1 & 0 & 1 & 4 & 5 & 11 \\ 
  NBPSS (MMR) & 3 & 0 & 0 & 0 & 2 & 100 & 0 & 79 & 2 & 0 & 8 & 6 & 0 & 0 & 3 & 0 & 1 & 2 & 6 & 11 \\ 
 \textbf{ NBPSS (DRB)} & 2 & 0 & 0 & 0 & 3 & 2 & 0 & 0 & 2 & 2 & 10 & 7 & 0 & 0 & 1 & 0 & 4 & 8 & 2 & \textbf{2} \\ 
\end{tabular}
\caption{Missclassification rates. Shown is the effectwise missclassification rate in percent across the $R=100$ replicates. Rows refer to methods while columns refer to effects. The first row indicates the effect (D stands for dummy, L for linear, N for nonlinear and S for spatial). The second row indicates whether the effect is truly active ($\gamma_j^\ast=1$) or not ($\gamma_j^\ast=0$). The rightmost column shows rowwise averages and can be regarded as an overall measure of selection accuracy.}
\label{tab:Missclassifcation}
\end{table}
The results are shown in Figure~\ref{fig:RMSE} and Table~\ref{tab:Missclassifcation}. 
Overall, the DR basis yields the best performance for both methods, NBPSS and \texttt{mboost}. Using the DR basis, NBPSS outperforms \texttt{mboost} in terms of both, RMSE and selection accuracy. Closer inspection of Table~\ref{tab:Missclassifcation} reveals that \texttt{mboost} tends to select overly complex models. This is in line with the literature and a current research topic \citep{MayWisSpe2023}. While it is not surprising that the DR basis leads to lower effectwise missclassification rates, the large gains in overall RMSE are somewhat surprising but consistent across different settings.
In Appendix~F.2 we consider several other scenarios, where we include a true spatial effect in the predictor, vary the sample size $n\in\{100,250,500,2000\}$ and introduce correlation among the continuous covariates $X_j$. 
The overall picture is well in line with the present results, and the combination of NBPSS and the DR basis consistently leads to the best performance.  

\subsection{Illustration: survival with leukemia}\label{sec:Application}

To illustrate the practical usefulness of the developed methodology, we analyze the \texttt{LeukSurv} data set. The data set is contained in the R package \texttt{spBayesSurv} \citep{ZhoHanZha2020} and has previously been analyzed by several different authors including \cite{HenShiGor2002,KneFah2007,LinRueLin2011,ZhoHan2018}. 

\paragraph{Model specification}
   The data set contains survival information for $n=1,043$ patients from Northwest England suffering from acute myeloid leukemia. About $16\%$ of the survival times are right-censored and available covariates are the patient's age (\textit{age}), the white blood cell count at diagnosis (\textit{wbc}), the Townsend deprivation index (\textit{tpi}) as well as the patient's sex (\textit{sex}). In addition to that, the exact residential location of the patients is available (\textit{space}). To contribute to a better understanding of the covariate effects, we consider the geoadditive Cox model
    \begin{align*}
       \log h(t_i\mid \bm{v}_i) =& g_0(t_i)+ \alpha sex_i+\beta_1 age_i+\beta_2 wbc_i+\beta_3 tpi_i\\&+f_{age}(age_i)+f_{wbc}(wbc_i)+f_{tpi}(tpi_i)+f_{spat}(space_i),\ i=1,\dots,n.
    \end{align*}
   We expand the nonlinear effects $f_{age},f_{wbc},f_{tpi}$ of the continuous covariates \textit{age}, \textit{wbc} and \textit{tpi} in terms of DR spline bases as explained in Section~\ref{sec:NewConstructionDRB}. For the spatial effect $f_{spat}$ we use tensor product B-splines and a Kronecker sum penalty. Then, we apply a DR type reparametrization to realize the centering constraints and to obtain a multiple of the identity matrix as penalty matrix (see Appendix~C.2 for details). For the log baseline hazard rate $g_0(\cdot)$ we use $d_0=10$ cubic Bayesian P-splines with second order differences penalty.  
    
\paragraph{MCMC sampling} We use $T=50,000$ MCMC iterations and discard the first $10,000$ as burn-in (overall runtime about 32 minutes). The MH acceptance rate for $\bm{\beta}_0$ was about $30\%$ and for $\bm{\beta}$ it was about $82\%$. Further MCMC diagnostics indicate that the sampler has converged to the desired target distribution (see Appendix~F.3).

\begin{figure}[htbp]
    \centering
    \includegraphics[width=\textwidth]{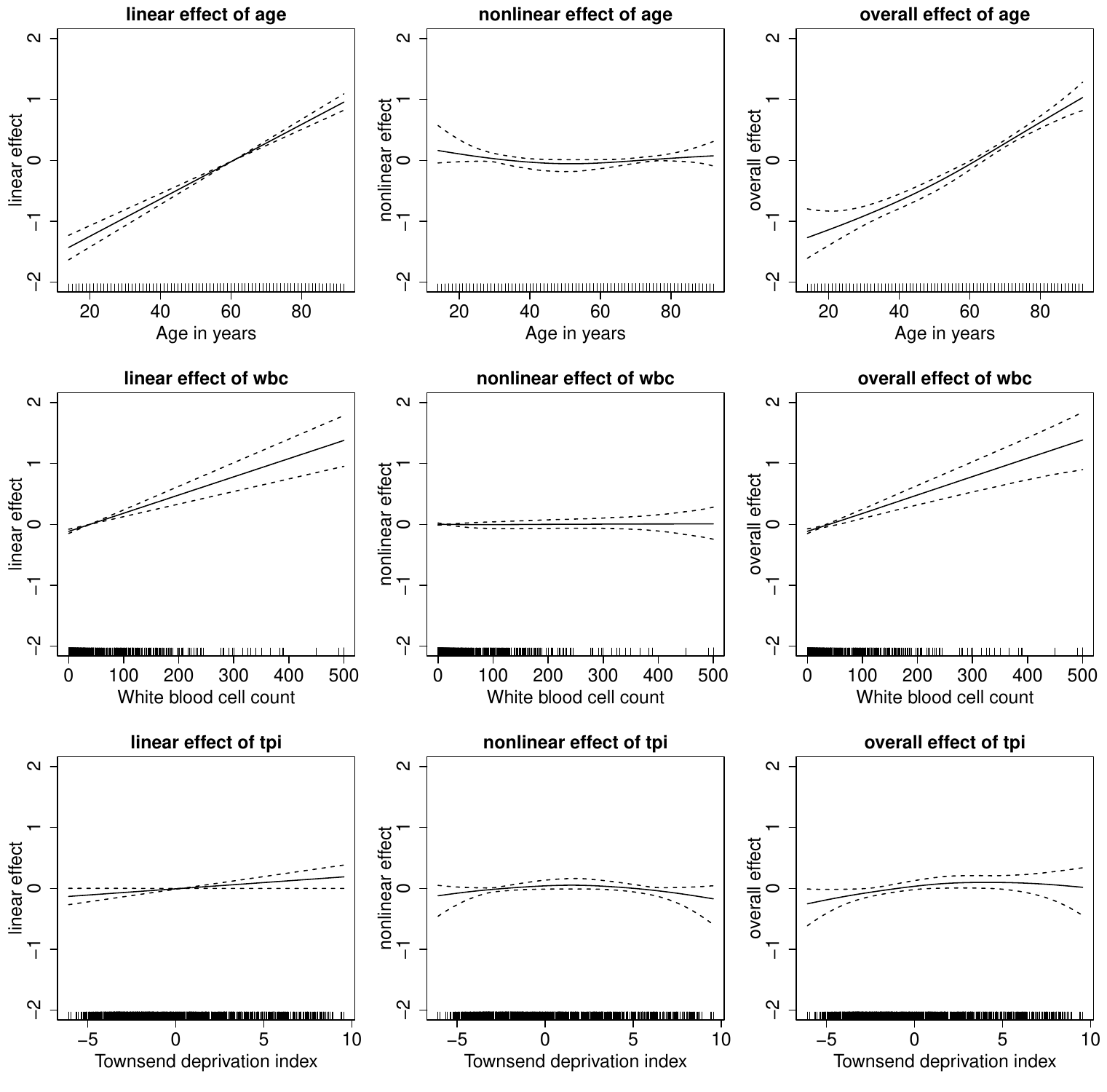}
    \caption{Survival with leukemia II. Estimated covariate effects.}
    \label{fig:LeukSurvCovariateEffects}
\end{figure}

\begin{figure}[htbp]
\centering
\begin{minipage}{.5\textwidth}
  \centering
\includegraphics[width=0.97\textwidth]{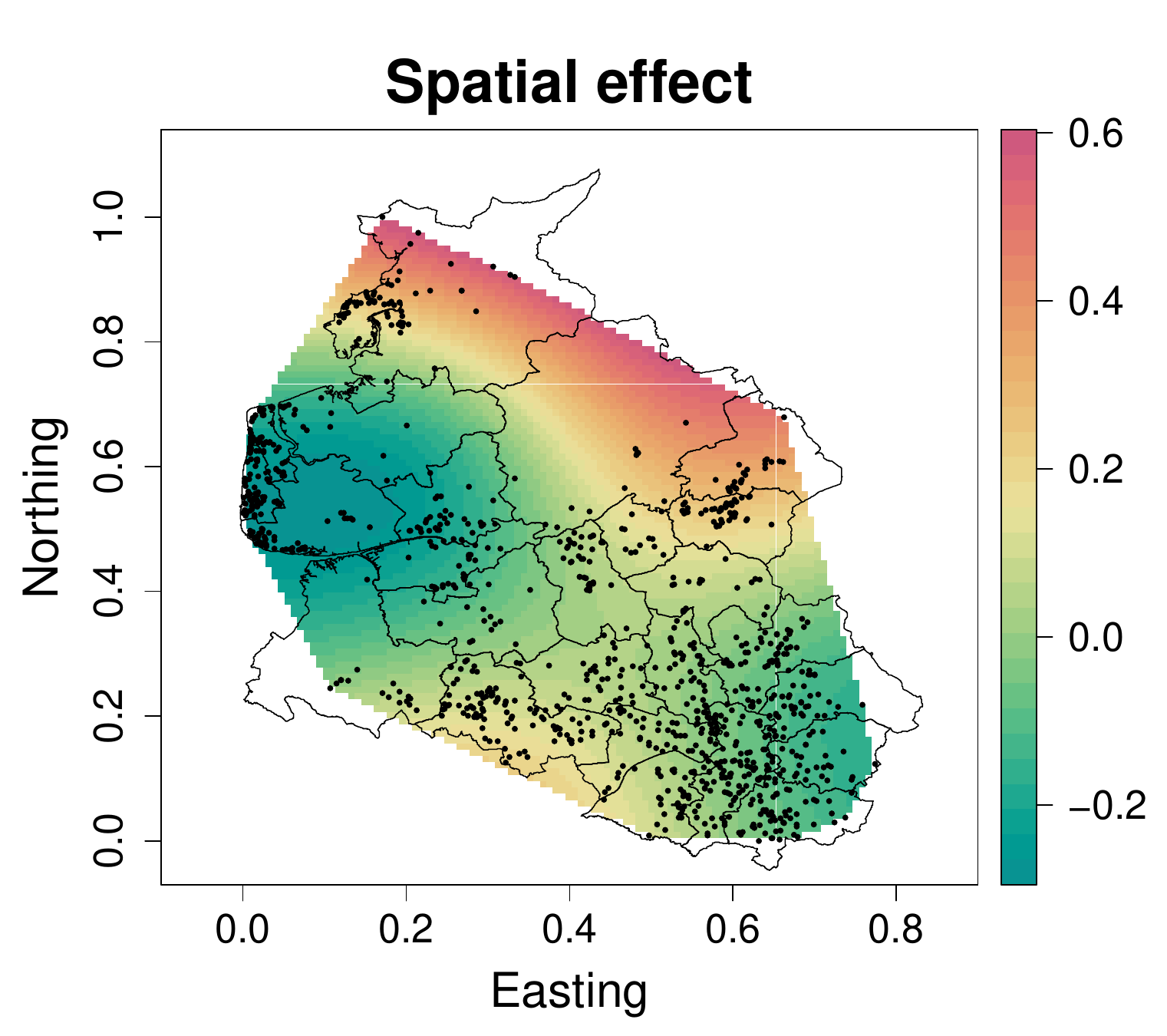}
\captionsetup{justification=centering}
    \captionof{figure}{Survival with leukemia III.\\ Estimated spatial effect.}
    \label{fig:LeukSurvSpatialEffect}
\end{minipage}%
\begin{minipage}{.5\textwidth}
  \centering
  \vspace{0.25em}
  \includegraphics[width=0.95\textwidth]{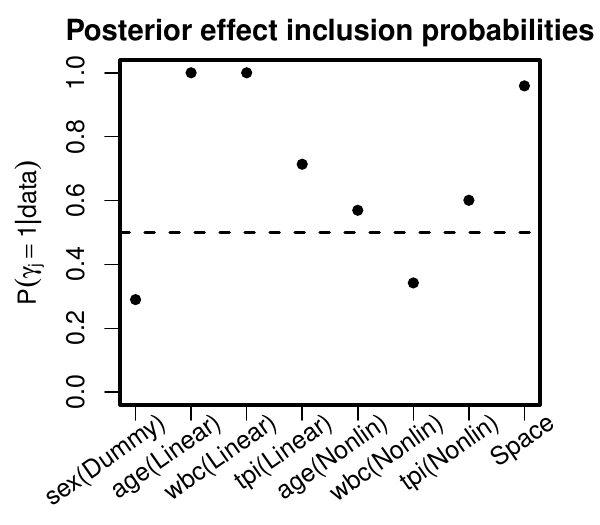}
  \captionsetup{justification=centering}
  \captionof{figure}{Survival with leukemia IV. Estimated posterior inclusion probabilities.}
  \label{fig:LeukSurvEstimatedProbabilities}
\end{minipage}
\end{figure}

    \paragraph{Results}
  Figure~\ref{fig:LeukSurvBaselineHazard} depicts the estimated log baseline hazard rate. We see that the hazard rate mainly decreases but sharply increases towards the end of the observation period. However, some caution is warranted as the observed time exceeds $10$ years for only $26$ patients. The estimated covariate and spatial effects are shown in Figures~\ref{fig:LeukSurvCovariateEffects} and~\ref{fig:LeukSurvSpatialEffect}, respectively.
Overall, the picture is well in line with the findings of \cite{KneFah2007}, who use an empirical Bayes approach for estimation and the Akaike information criterion (AIC) for model selection. Interestingly, \cite{KneFah2007} fit a range of different models and conclude that the fully-nonparametric model with nonlinear effects for all continuous covariates performs best in terms of AIC, even though the overall effect of \textit{wbc} appears to be linear from their plot too (as in our case). Using the posterior inclusion probabilities, we conclude that a linear effect should actually be sufficient for modeling \textit{wbc}. We acknowledge, however, that the inclusion probability is fairly close to the threshold of $0.5$  (see Figure~\ref{fig:LeukSurvEstimatedProbabilities}). 

\begin{figure}[htbp]
    \centering
    \includegraphics[width=0.7\textwidth]{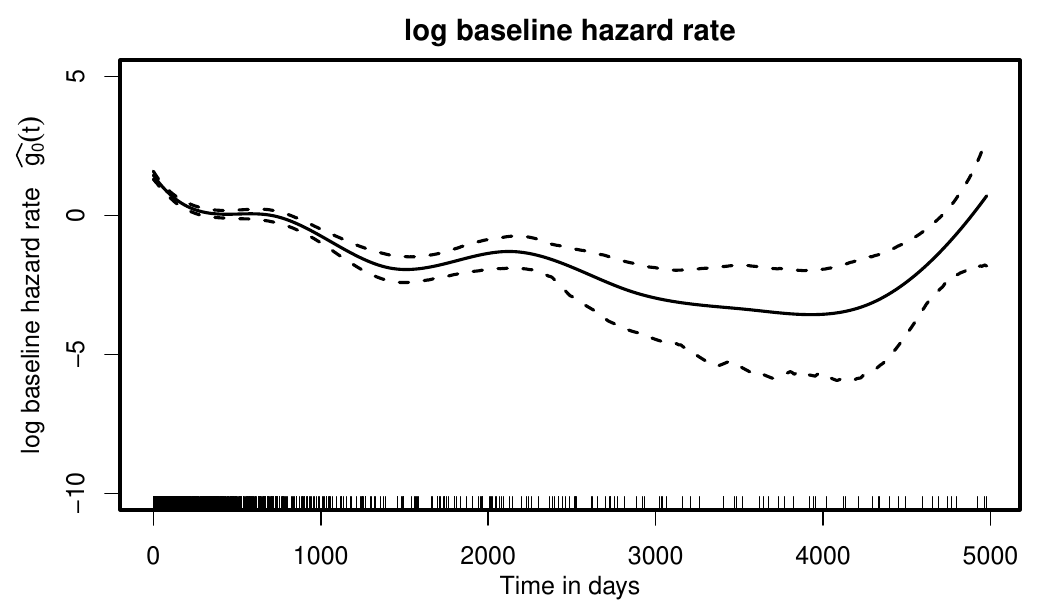}
    \caption{Survival with leukemia I. Estimated log-baseline hazard rate.}
    \label{fig:LeukSurvBaselineHazard}
\end{figure}

\section{Discussion and Conclusion}\label{sec:Discussion}
The main contribution of this paper is to highlight the importance of orthogonality between the linear and the nonlinear effect for  efficiently estimating sparse partially linear additive models. Only then do we obtain consistent estimation of the true linear and nonlinear effect, which we define in terms of $\mathcal{L}^2(\dsP^{X_j})$-projections. To achieve orthogonality, we expand the nonlinear effects in DR spline bases. We suggest a new construction of the DR basis for P-splines, which is more efficient and less restrictive than previous suggestions as it does not require full rank of the B-spline design matrix. Thereby, our proposal can be seen as an attractive default for additive P-spline regression, irrespective of whether a Bayesian,  boosting or  penalized likelihood approach is used for estimation. Our simulations for the geoadditive Cox model show that the DR basis is not only theoretically but also empirically superior to the commonly used MMRs. Interestingly, not only our Bayesian method but also the boosting approach of \cite{HofMayRob2014} benefits from using the DR basis for the nonlinear effects.

Another important contribution of this work is that we allow for a better understanding of the superior MCMC mixing properties of the recently proposed NBPSS prior. To this end, we analytically derive and investigate the implied spike and slab of the Euclidean norm of the group coefficient vectors. In general, we argue that the implied prior density of the norm is a key quantity for Bayesian group selection approaches and should always be investigated and visualized. 

Two promising directions for future research are as follows. First, as an alternative to the geoadditive Cox model, one may investigate Bayesian effect selection for parametric survival models such as a geoadditive Weibull regression model or a Weibull model with geoadditive predictors on both, the scale and the shape parameter. We have focused on the Cox model because it is still the most popular survival model in applications \citep{GeoSeaAba2014}, but the investigation of parametric alternatives would be very interesting. Second, as a potential alternative to MCMC, one may investigate Bayesian effect selection for non-standard likelihoods using faster approximate Bayesian methods such as variational inference. \cite{HeWan2023} provide such an option using a mean field approximation but their approach is limited to a Gaussian response model and a binary probit model, and it would definitely be interesting to broaden the scope in future research. For both of these directions it will be important to have an orthogonal decomposition and the suggested construction of the DR basis will be convenient to model the nonlinear effects of continuous covariates.

\section*{Acknowledgements}

This work was funded by the Deutsche Forschungsgemeinschaft (DFG, German Research Foundation) through the Emmy Noether grant KL 3037/1-1. 

\bibliographystyle{apalike}
\bibliography{sample}

\begin{thebibliography}{}

\bibitem[Abramowitz and Stegun, 1972]{AbrSte1972}
Abramowitz, M. and Stegun, I.~A. (1972).
\newblock {\em Handbook of Mathematical Functions with Formulas, Graphs, and Mathematical Tables}.
\newblock US Government Printing Office, Washington, D.C.
\newblock tenth printing.

\bibitem[Aguilar and B{\"u}rkner, 2023]{AguBur2023}
Aguilar, J.~E. and B{\"u}rkner, P.-C. (2023).
\newblock Intuitive joint priors for {B}ayesian linear multilevel models: {T}he {R2D2M2} prior.
\newblock {\em Electronic Journal of Statistics}, 17(1):1711--1767.

\bibitem[Armagan et~al., 2013]{ArmDunLee2013b}
Armagan, A., Dunson, D.~B., Lee, J., Bajwa, W.~U., and Strawn, N. (2013).
\newblock Posterior consistency in linear models under shrinkage priors.
\newblock {\em Biometrika}, 100(4):1011--1018.

\bibitem[Bai et~al., 2022]{BaiMorAnt2022}
Bai, R., Moran, G.~E., Antonelli, J.~L., Chen, Y., and Boland, M.~R. (2022).
\newblock Spike-and-slab group lassos for grouped regression and sparse generalized additive models.
\newblock {\em Journal of the American Statistical Association}, 117(537):184--197.

\bibitem[Barbieri and Berger, 2004]{BarBer2004}
Barbieri, M.~M. and Berger, J.~O. (2004).
\newblock Optimal predictive model selection.
\newblock {\em The Annals of Statistics}, 32(3):870--897.

\bibitem[Barbieri et~al., 2021]{BarBerGeo2021}
Barbieri, M.~M., Berger, J.~O., George, E.~I., and Ro{\v{c}}kov{\'a}, V. (2021).
\newblock The median probability model and correlated variables.
\newblock {\em Bayesian Analysis}, 16(4):1085--1112.

\bibitem[Berk, 2008]{Ber2008}
Berk, R.~A. (2008).
\newblock {\em Statistical Learning from a Regression Perspective}.
\newblock Springer Nature, Cham, Switzerland, third edition.

\bibitem[Brilleman et~al., 2020]{BriElcNov2020}
Brilleman, S.~L., Elci, E.~M., Novik, J.~B., and Wolfe, R. (2020).
\newblock Bayesian survival analysis using the rstanarm {R} package.
\newblock {\em arXiv preprint arXiv:2002.09633}.

\bibitem[Chouldechova and Hastie, 2015]{ChoHas2015}
Chouldechova, A. and Hastie, T. (2015).
\newblock Generalized additive model selection.
\newblock {\em arXiv preprint arXiv:1506.03850}.

\bibitem[Claeskens et~al., 2009]{ClaKriOps2009}
Claeskens, G., Krivobokova, T., and Opsomer, J.~D. (2009).
\newblock Asymptotic properties of penalized spline estimators.
\newblock {\em Biometrika}, 96(3):529--544.

\bibitem[Cox, 1972]{Cox1972b}
Cox, D.~R. (1972).
\newblock Regression models and life-tables.
\newblock {\em Journal of the Royal Statistical Society: Series B (Methodological)}, 34(2):187--220.

\bibitem[de~Boor, 2001]{Boo2001}
de~Boor, C. (2001).
\newblock {\em A Practical Guide to Splines}.
\newblock Springer, New York, NY, revised edition.

\bibitem[Demmler and Reinsch, 1975]{DemRei1975}
Demmler, A. and Reinsch, C. (1975).
\newblock Oscillation matrices with spline smoothing.
\newblock {\em Numerische Mathematik}, 24(5):375--382.

\bibitem[Eilers, 1999]{Eil1999}
Eilers, P. H.~C. (1999).
\newblock Comment on ``{T}he analysis of designed experiments and longitudinal data by using smoothing splines'' by {V}erbyla, {A}. {P}., et al.
\newblock {\em Journal of the Royal Statistical Society: Series C (Applied Statistics)}, 48(3):307--308.

\bibitem[Eilers and Marx, 1996]{EilMar1996}
Eilers, P. H.~C. and Marx, B.~D. (1996).
\newblock Flexible smoothing with {B}-splines and penalties.
\newblock {\em Statistical Science}, 11(2):89--121.

\bibitem[Fahrmeir et~al., 2004]{FahKneLan2004}
Fahrmeir, L., Kneib, T., and Lang, S. (2004).
\newblock Penalized structured additive regression for space-time data: a {B}ayesian perspective.
\newblock {\em Statistica Sinica}, 14(3):731--761.

\bibitem[Fahrmeir et~al., 2021]{FahKneLan2021}
Fahrmeir, L., Kneib, T., Lang, S., and Marx, B.~D. (2021).
\newblock {\em Regression: Models, Methods and Applications}.
\newblock Springer, Berlin, second edition.

\bibitem[Gamerman, 1997]{Gam1997}
Gamerman, D. (1997).
\newblock Sampling from the posterior distribution in generalized linear mixed models.
\newblock {\em Statistics and Computing}, 7(1):57--68.

\bibitem[George et~al., 2014]{GeoSeaAba2014}
George, B., Seals, S., and Aban, I. (2014).
\newblock Survival analysis and regression models.
\newblock {\em Journal of Nuclear Cardiology}, 21(4):686--694.

\bibitem[George and McCulloch, 1993]{GeoMcC1993}
George, E.~I. and McCulloch, R.~E. (1993).
\newblock Variable selection via {G}ibbs sampling.
\newblock {\em Journal of the American Statistical Association}, 88(423):881--889.

\bibitem[George and McCulloch, 1997]{GeoMcC1997}
George, E.~I. and McCulloch, R.~E. (1997).
\newblock Approaches for {B}ayesian variable selection.
\newblock {\em Statistica Sinica}, 7(2):339--373.

\bibitem[Guo et~al., 2022]{GuoJaeRah2022}
Guo, B., Jaeger, B.~C., Rahman, A.~F., Long, D.~L., and Yi, N. (2022).
\newblock Spike-and-slab least absolute shrinkage and selection operator generalized additive models and scalable algorithms for high--dimensional data analysis.
\newblock {\em Statistics in Medicine}, 41(20):3899--3914.

\bibitem[Hastie and Tibshirani, 1986]{HasTib1986}
Hastie, T. and Tibshirani, R. (1986).
\newblock Generalized additive models.
\newblock {\em Statistical Science}, 1(3):297--310.

\bibitem[Hastie and Tibshirani, 1990]{HasTib1990}
Hastie, T. and Tibshirani, R. (1990).
\newblock {\em Generalized Additive Models}.
\newblock {Chapman and Hall/CRC Press}, Boca Raton, FL.

\bibitem[He and Wand, 2023]{HeWan2023}
He, V.~X. and Wand, M.~P. (2023).
\newblock Bayesian generalized additive model selection including a fast variational option.
\newblock {\em AStA Advances in Statistical Analysis}.

\bibitem[Henderson et~al., 2002]{HenShiGor2002}
Henderson, R., Shimakura, S., and Gorst, D. (2002).
\newblock Modeling spatial variation in leukemia survival data.
\newblock {\em Journal of the American Statistical Association}, 97(460):965--972.

\bibitem[Hennerfeind et~al., 2006]{HenBreFah2006}
Hennerfeind, A., Brezger, A., and Fahrmeir, L. (2006).
\newblock Geoadditive survival models.
\newblock {\em Journal of the American Statistical Association}, 101(475):1065--1075.

\bibitem[Hern{\'a}ndez-Lobato et~al., 2013]{HerHerDup2013}
Hern{\'a}ndez-Lobato, D., Hern{\'a}ndez-Lobato, J.~M., and Dupont, P. (2013).
\newblock Generalized spike-and-slab priors for {B}ayesian group feature selection using expectation propagation.
\newblock {\em Journal of Machine Learning Research}, 14(59):1891--1945.

\bibitem[Hofner et~al., 2014]{HofMayRob2014}
Hofner, B., Mayr, A., Robinzonov, N., and Schmid, M. (2014).
\newblock Model-based boosting in {R}: a hands-on tutorial using the {R} package mboost.
\newblock {\em Computational Statistics}, 29(1-2):3--35.

\bibitem[Hu et~al., 2015]{HuZhaLia2015}
Hu, Y., Zhao, K., and Lian, H. (2015).
\newblock Bayesian quantile regression for partially linear additive models.
\newblock {\em Statistics and Computing}, 25(3):651--668.

\bibitem[Huang, 1998]{Hua1998b}
Huang, J.~Z. (1998).
\newblock Projection estimation in multiple regression with application to functional {ANOVA} models.
\newblock {\em The Annals of Statistics}, 26(1):242--272.

\bibitem[Kelker, 1970]{Kel1970}
Kelker, D. (1970).
\newblock Distribution theory of spherical distributions and a location-scale parameter generalization.
\newblock {\em Sankhyā: The Indian Journal of Statistics, Series A}, 32(4):419--430.

\bibitem[Klein et~al., 2021]{KleCarKne2021}
Klein, N., Carlan, M., Kneib, T., Lang, S., and Wagner, H. (2021).
\newblock Bayesian effect selection in structured additive distributional regression models.
\newblock {\em Bayesian Analysis}, 16(2):545--573.

\bibitem[Kneib and Fahrmeir, 2007]{KneFah2007}
Kneib, T. and Fahrmeir, L. (2007).
\newblock A mixed model approach for geoadditive hazard regression.
\newblock {\em Scandinavian Journal of Statistics}, 34(1):207--228.

\bibitem[Kyung et~al., 2010]{KyuGilGho2010}
Kyung, M., Gill, J., Ghosh, M., and Casella, G. (2010).
\newblock Penalized regression, standard errors, and {B}ayesian lassos.
\newblock {\em Bayesian Analysis}, 5(2):369--411.

\bibitem[Lang and Brezger, 2004]{LanBre2004}
Lang, S. and Brezger, A. (2004).
\newblock Bayesian {P}-splines.
\newblock {\em Journal of Computational and Graphical Statistics}, 13(1):183--212.

\bibitem[Lindgren et~al., 2011]{LinRueLin2011}
Lindgren, F., Rue, H., and Lindstr{\"o}m, J. (2011).
\newblock An explicit link between {G}aussian fields and {G}aussian {M}arkov random fields: the stochastic partial differential equation approach.
\newblock {\em Journal of the Royal Statistical Society: Series B (Statistical Methodology)}, 73(4):423--498.

\bibitem[Lou et~al., 2016]{LouBieCar2016}
Lou, Y., Bien, J., Caruana, R., and Gehrke, J. (2016).
\newblock Sparse partially linear additive models.
\newblock {\em Journal of Computational and Graphical Statistics}, 25(4):1126--1140.

\bibitem[Martino et~al., 2011]{MarAkeRue2011}
Martino, S., Akerkar, R., and Rue, H. (2011).
\newblock Approximate {B}ayesian inference for survival models.
\newblock {\em Scandinavian Journal of Statistics}, 38(3):514--528.

\bibitem[Mayr et~al., 2023]{MayWisSpe2023}
Mayr, A., Wistuba, T., Speller, J., Gude, F., and Hofner, B. (2023).
\newblock Linear or smooth? {E}nhanced model choice in boosting via deselection of base-learners.
\newblock {\em Statistical Modelling}, 23(5-6):441--455.

\bibitem[Neal, 2003]{Nea2003}
Neal, R.~M. (2003).
\newblock Slice sampling.
\newblock {\em The Annals of Statistics}, 31(3):705--767.

\bibitem[Nychka and Cummins, 1996]{NycCum1996}
Nychka, D. and Cummins, D. (1996).
\newblock Comment on ``{F}lexible smoothing with {B}-splines and penalties'' by {E}ilers, {P}. {H}. {C}. and {M}arx, {B}. {D}.
\newblock {\em Statistical Science}, 11(2):104--105.

\bibitem[O'Sullivan, 1986]{OS1986}
O'Sullivan, F. (1986).
\newblock A statistical perspective on ill-posed inverse problems.
\newblock {\em Statistical Science}, 1(4):502--518.

\bibitem[Parlett, 1998]{Par1998}
Parlett, B.~N. (1998).
\newblock {\em The Symmetric Eigenvalue Problem}.
\newblock SIAM, Philadelphia, PA.

\bibitem[P{\'e}rez et~al., 2017]{PerPerRam2017}
P{\'e}rez, M.-E., Pericchi, L.~R., and Ram{\'i}rez, I.~C. (2017).
\newblock The scaled beta2 distribution as a robust prior for scales.
\newblock {\em Bayesian Analysis}, 12(3):615--637.

\bibitem[Ravikumar et~al., 2009]{RavLafLiu2009}
Ravikumar, P., Lafferty, J., Liu, H., and Wasserman, L. (2009).
\newblock Sparse additive models.
\newblock {\em Journal of the Royal Statistical Society: Series B (Statistical Methodology)}, 71(5):1009--1030.

\bibitem[Robert and Roberts, 2021]{RobRob2021}
Robert, C.~P. and Roberts, G. (2021).
\newblock Rao--{B}lackwellisation in the {M}arkov chain {M}onte {C}arlo era.
\newblock {\em International Statistical Review}, 89(2):237--249.

\bibitem[Ro{\v{c}}kov{\'a}, 2018]{Roc2018}
Ro{\v{c}}kov{\'a}, V. (2018).
\newblock Bayesian estimation of sparse signals with a continuous spike-and-slab prior.
\newblock {\em The Annals of Statistics}, 46(1):401--437.

\bibitem[Rossell and Rubio, 2023]{RosRub2023}
Rossell, D. and Rubio, F.~J. (2023).
\newblock Additive {B}ayesian variable selection under censoring and misspecification.
\newblock {\em Statistical Science}, 38(1):13--29.

\bibitem[Ruppert, 2002]{Rup2002}
Ruppert, D. (2002).
\newblock Selecting the number of knots for penalized splines.
\newblock {\em Journal of Computational and Graphical Statistics}, 11(4):735--757.

\bibitem[Scheipl et~al., 2012]{SchFahKne2012}
Scheipl, F., Fahrmeir, L., and Kneib, T. (2012).
\newblock Spike-and-slab priors for function selection in structured additive regression models.
\newblock {\em Journal of the American Statistical Association}, 107(500):1518--1532.

\bibitem[Schoenberg and Whitney, 1953]{SchWhi1953}
Schoenberg, I.~J. and Whitney, A. (1953).
\newblock On {P\'o}lya frequency functions. {III}. {T}he positivity of translation determinants with an application to the interpolation problem by spline curves.
\newblock {\em Transactions of the American Mathematical Society}, 74(2):246--259.

\bibitem[Speckman, 1985]{Spe1985}
Speckman, P. (1985).
\newblock Spline smoothing and optimal rates of convergence in nonparametric regression models.
\newblock {\em The Annals of Statistics}, 13(3):970--983.

\bibitem[Stone, 1986]{Sto1986}
Stone, C.~J. (1986).
\newblock The dimensionality reduction principle for generalized additive models.
\newblock {\em The Annals of Statistics}, 14(2):590--606.

\bibitem[Stone, 1994]{Sto1994}
Stone, C.~J. (1994).
\newblock The use of polynomial splines and their tensor products in multivariate function estimation.
\newblock {\em The Annals of Statistics}, 22(1):118--171.

\bibitem[{van de Geer}, 2000]{Gee2000}
{van de Geer}, S.~A. (2000).
\newblock {\em Empirical Processes in M-estimation}.
\newblock {Cambridge University Press}, Cambridge, UK.

\bibitem[Vershynin, 2018]{Ver2018}
Vershynin, R. (2018).
\newblock {\em High-Dimensional Probability: An Introduction with Applications in Data Science}.
\newblock {Cambridge University Press}, Cambridge, UK.

\bibitem[Wahba, 1990]{Wah1990}
Wahba, G. (1990).
\newblock {\em Spline Models for Observational Data}.
\newblock SIAM, Philadelphia, PA.

\bibitem[Wand and Ormerod, 2008]{WanOrm2008}
Wand, M.~P. and Ormerod, J.~T. (2008).
\newblock On semiparametric regression with {O'Sullivan} penalized splines.
\newblock {\em Australian {\&} New Zealand Journal of Statistics}, 50(2):179--198.

\bibitem[Wiemann et~al., 2021]{WieKneWag2021b}
Wiemann, P., Kneib, T., and Wagner, H. (2021).
\newblock Effect selection and regularization in structured additive distributional regression.
\newblock In {\em Handbook of Bayesian Variable Selection}, pages 271--296. {Chapman and Hall/CRC Press}, Boca Raton, FL.

\bibitem[Wood, 2017]{Woo2017b}
Wood, S.~N. (2017).
\newblock {\em Generalized Additive Models: An Introduction with R}.
\newblock {Chapman and Hall/CRC Press}, Boca Raton, FL, second edition.

\bibitem[Yuan and Lin, 2006]{YuaLin2006}
Yuan, M. and Lin, Y. (2006).
\newblock Model selection and estimation in regression with grouped variables.
\newblock {\em Journal of the Royal Statistical Society: Series B (Statistical Methodology)}, 68(1):49--67.

\bibitem[Zhang et~al., 2011]{ZhaCheLiu2011}
Zhang, H.~H., Cheng, G., and Liu, Y. (2011).
\newblock Linear or nonlinear? {A}utomatic structure discovery for partially linear models.
\newblock {\em Journal of the American Statistical Association}, 106(495):1099--1112.

\bibitem[Zhou and Hanson, 2018]{ZhoHan2018}
Zhou, H. and Hanson, T. (2018).
\newblock A unified framework for fitting {B}ayesian semiparametric models to arbitrarily censored survival data, including spatially referenced data.
\newblock {\em Journal of the American Statistical Association}, 113(522):571--581.

\bibitem[Zhou et~al., 2020]{ZhoHanZha2020}
Zhou, H., Hanson, T., and Zhang, J. (2020).
\newblock {spBayesSurv}: Fitting {B}ayesian spatial survival models using {R}.
\newblock {\em Journal of Statistical Software}, 92(9):1--33.

\end{thebibliography}

\end{document}